\documentclass[%
 reprint,
preprint,
onecolumn,11pt,
amsmath,amssymb,aps,
prb,
]{revtex4-1}

\usepackage{verbatim}
\usepackage{graphicx}
\usepackage{physics}
\normalfont \topmargin -0.5 cm
\begin{document}

\title{Measurement of the spin temperature of optically cooled nuclei and GaAs hyperfine constants in GaAs/AlGaAs quantum dots}

\author{E. A. Chekhovich$^{1}$}\email{e.chekhovich@sheffield.ac.uk}
\author{A. Ulhaq$^1$}
\author{E. Zallo$^{2,3}$}
\author{F. Ding$^2$}
\author{O. G. Schmidt$^2$}
\author{M. S. Skolnick$^1$}
\affiliation{$^1$Department of Physics and Astronomy, University
of Sheffield, Sheffield, S3 7RH, United Kingdom.}
\affiliation{$^2$Institute for Integrative Nanoscience, IFW
Dresden, Helmholtz str. D-01069, Dresden, Germany.}
\affiliation{$^3$Paul-Drude-Institut f\"{u}r
Festk\"{o}rperelektronik, Hausvogteiplatz 5-7, 10117 Berlin,
Germany.}



\maketitle


\textbf{
Deep cooling of electron and nuclear spins is equivalent to
achieving polarization degrees close to 100\% and is a key
requirement in solid state quantum information technologies
\cite{AtatureScience2006,PhysRevLett.102.057403,PhysRevLett.114.247603,PlaNature2013,BurkardPRL2009,PhysRevB.66.245303}.
While polarization of individual nuclear spins in diamond
\cite{PhysRevLett.102.057403} and SiC
\cite{PhysRevLett.114.247603} reaches 99\% and beyond, it has been
limited to 60-65\% for the nuclei in quantum dots
\cite{GammonPRL2001,ChekhovichPRL2010}. Theoretical models have
attributed this limit to formation of coherent ''dark'' nuclear
spin states
\cite{ImamogluPRL2003,PhysRevB.75.155324,HildmannPRB2014} but
experimental verification is lacking, especially due to the poor
accuracy of polarization degree measurements. Here we measure the
nuclear polarization in GaAs/AlGaAs quantum dots with high
accuracy using a new approach enabled by manipulation of the
nuclear spin states with radiofrequency pulses. Polarizations up
to 80\% are observed -- the highest reported so far for optical
cooling in quantum dots. This value is still not limited by
nuclear coherence effects. Instead we find that optically cooled
nuclei are well described within a classical spin temperature
framework \cite{GoldmanBook}. Our findings unlock a route for
further progress towards quantum dot electron spin qubits where
deep cooling of the mesoscopic nuclear spin ensemble is used to
achieve long qubit coherence
\cite{PhysRevB.66.245303,BurkardPRL2009}. Moreover, GaAs hyperfine
material constants are measured here experimentally for the first
time.}

Optically active III-V semiconductors quantum dots are considered
for applications in quantum information technologies and have
major advantages, such as versatile device fabrication techniques
and strong interaction between charge spin and light
\cite{PhysRevLett.116.020401}. However, magnetic coupling with the
randomly polarized nuclei of the quantum dot, makes the spin state
of the electron vulnerable to dephasing and decoherence
\cite{Bechtold2015,PressEcho}. Improvement of the electron spin
coherence by preparing the nuclear spin ensemble in a ''narrowed''
state \cite{PhysRevB.66.245303,BurkardPRL2009,BarnesPRL2012} has
been demonstrated experimentally
\cite{Latta2009,BluhmPRL2010,SunPRL2012}. However, the actual
microscopic nature of the nuclear spin bath state remains unknown.
Moreover, the ultimately ''narrowed'', fully polarized nuclear
spin state not only has not been achieved, but it has been
difficult to identify the obstacles to such 100\% polarization.
The complexity of the problem arises from the mesoscopic nature of
the nuclear spin bath of a quantum dot: the typical number of
spins $10^4-10^6$ is too large to access each individual nucleus,
yet too small to ignore quantum correlations, coherence and
fluctuations
\cite{ImamogluPRL2003,PhysRevB.75.155324,HildmannPRB2014,BurkardPRL2009}.
The problem is complicated further by inhomogeneity of the
electron-nuclear interaction within the quantum dot volume.

In quantum wells and dots the degree of nuclear spin polarization
$P_\textrm{N}$ was previously estimated
\cite{PhysRevB.56.4743,GammonPRL2001} by measuring the resulting
hyperfine shift $E_\textrm{hf}$, which is the change in the energy
splitting of the $S_\textrm{z}=\pm$1/2 electron spin levels. The
shift produced by each nuclear isotope is:
\begin{equation}
E_\textrm{hf} = k A I P_\textrm{N},\label{eq:Ehf}
\end{equation}
where $I$ is the nuclear spin, $A$ is the hyperfine constant
characterizing the isotope and material only, and $k$ is a factor
($0\leq k\leq1$) describing the spatial non-uniformities of the
nuclear polarization, electron envelope wavefunction, and chemical
composition in a specific quantum dot structure. While
$E_\textrm{hf}$ can be measured very accurately, the uncertainty
in $A$ and $k$, leads to uncertainty in $P_\textrm{N}$. Here we
demonstrate measurement of $P_\textrm{N}$ not relying on
assumptions about $A$ and $k$, but based on direct mapping of the
spin-3/2 eigenstate populations. We then estimate $k$ from the
first-principles calculations and measurements on quantum dots of
different size which allows Eq. \ref{eq:Ehf} to be factorised and
the hyperfine constants $A$ to be derived accurately. Notably,
despite GaAs being one of the most important semiconductors, its
hyperfine material constants have not yet been measured
experimentally, but only estimated \cite{Paget1977} using the
hyperfine constants of InSb obtained more than five decades ago
\cite{Gueron}.

We achieve $P_\textrm{N}\approx80$\% corresponding to spin
temperature $T_\textrm{N}\approx1.3$~mK at a bath temperature of
$T=4.2$~K. The observed $P_\textrm{N}$ exceeds the predicted
values for the quantum limit for nuclear spin cooling
\cite{PhysRevB.75.155324,HildmannPRB2014}. Furthermore, we have
implemented the proposed protocols where electron-nuclear coupling
is ''modulated'' during optical cooling
\cite{ImamogluPRL2003,PhysRevB.75.155324} but observed no increase
in $P_\textrm{N}$, ruling out coherent ''dark'' nuclear spin
states as a single fundamental obstacle. Instead we expect the
currently achieved $P_\textrm{N}$ to be limited by competing
contributions of nuclear spin pumping and depolarization
mechanisms -- with a further effort in designing the nuclear spin
cooling protocol these obstacles can be overcome, potentially
opening the way for achieving nuclear polarizations close to
100\%.

\begin{figure}[h]
\includegraphics[bb=50pt 150pt 630pt 730pt, width=11cm]{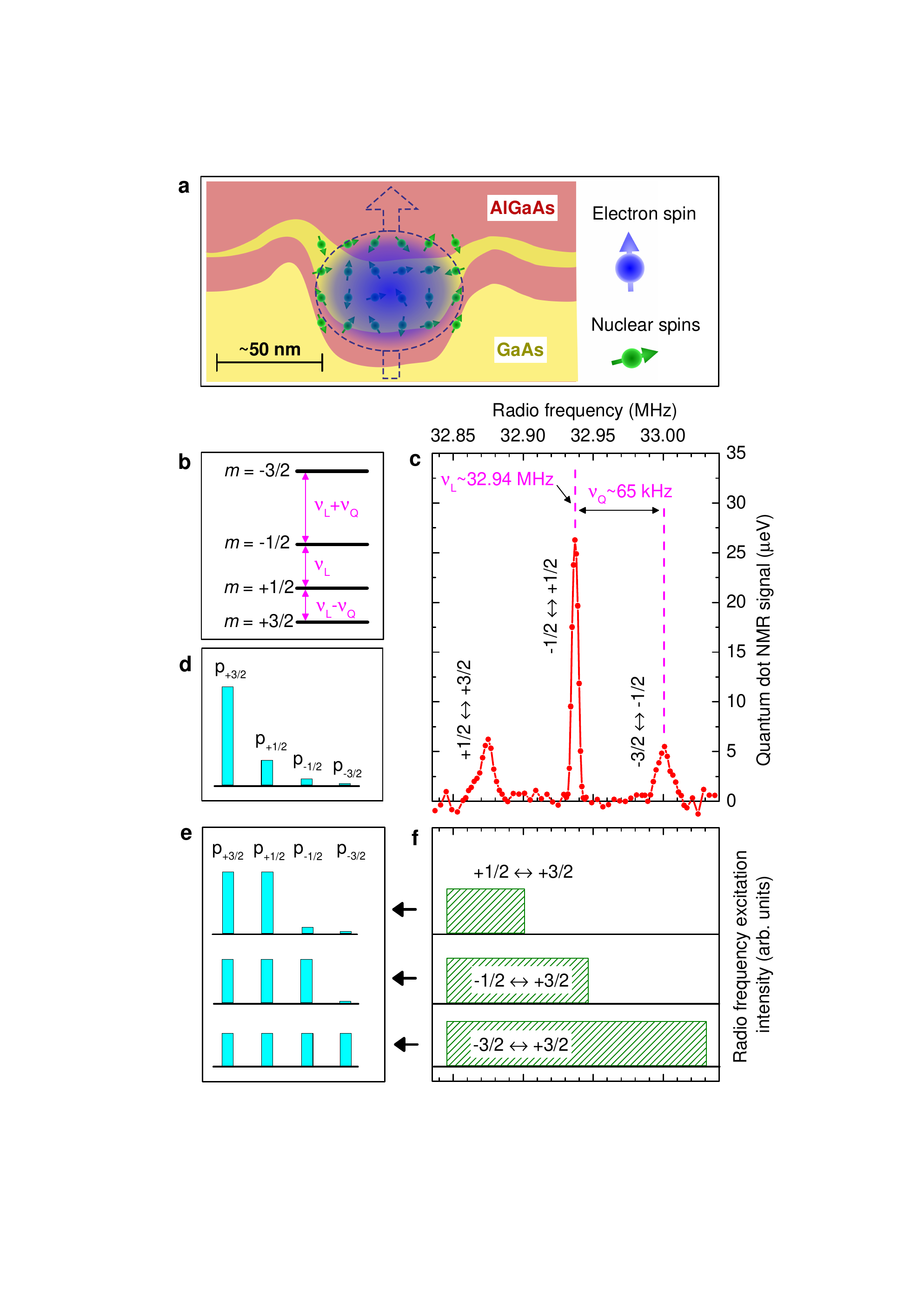}
\caption{\textbf{Manipulation of the nuclear spin states in
quantum dots.} \textbf{a,} Schematic of a nanohole in-filled
GaAs/AlGaAs quantum dot. An electron trapped in the dot interacts
with $>10^4$ nuclear spins. \textbf{b,} Energy levels of a nuclear
spin $I$=3/2 have distinct spin projections $m$ along magnetic
field $B_\textrm{z}$. Level splitting is dominated by the Zeeman
effect (characterized by Larmor frequency $\nu_\textrm{L}\propto
B_\textrm{z}$), qadrupolar effects cause additional changes in
energies (characterized by $\nu_\textrm{Q}\ll \nu_\textrm{L}$).
\textbf{(c,)} Nuclear magnetic resonance (NMR) spectrum measured
on $^{75}$As nuclei of a single quantum dot B1 at
$B_z\approx$4.5~T using ''inverse'' NMR method
\cite{ChekhovichNatNano2012}: transitions between individual
levels $m$ are clearly resolved due to strain induced quadrupolar
effects. \textbf{d,} Schematic of initial population probabilities
$p_{m}$ of the nuclear spin levels with different $m$
corresponding to Boltzmann distribution with polarization degree
$P_N\approx+$80\%. \textbf{e,f,} Modified populations of the
nuclear spin levels (\textbf{e}) resulting from selective
saturation of the nuclear spin transitions with radio frequency
magnetic field with rectangular shaped spectra shown in
(\textbf{f}).} \label{fig1}
\end{figure}

We study GaAs/AlGaAs quantum dots (QDs) grown by in-situ nanohole
etching and filling \cite{Atkinson2012}. A schematic cross-section
of such a QD is shown in Fig.~\ref{fig1}a. An individual electron
with spin $S$=1/2 (blue) can be trapped in a QD typically
consisting of $\sim10^5$ atoms (green), each possessing a nuclear
spin $I$=3/2 for gallium and arsenic, or $I$=5/2 for aluminium.
The electron-nuclear magnetic interaction (known as the hyperfine
interaction) has a dual effect. Firstly, spin polarized electrons
injected repeatedly into a QD via optical excitation can exchange
spin with the nuclei (flip-flop process) resulting in nuclear spin
cooling (also referred to as dynamic nuclear polarization, or DNP)
\cite{GammonPRL2001,ChekhovichPRL2010}. Secondly, the net nuclear
polarization produces a hyperfine shift $E_\textrm{hf}$ in the
electron spin state energy splitting. Such shifts are detected
optically in the luminescence spectrum of a QD and are used to
probe the nuclear polarization
\cite{GammonPRL2001,ChekhovichPRL2010}. (Further details on
samples and experimental techniques can be found in Methods and
Supplementary Notes 1 and 2.)

All experiments are performed at $T$=4.2~K and magnetic field
$B_\textrm{z}>4$~T along the sample growth axis ($z$). Under such
conditions the nuclear spin states have well-defined projections
$m$ along the $z$-axis. The structure of the nuclear spin energy
levels is sketched in Fig.~\ref{fig1}b: magnetic field induces
Zeeman shifts $mh\nu_\textrm{L}$ ($h$ is Planck's constant), so
that the frequencies of all dipole-allowed transitions $m
\leftrightarrow m+1$ equal the Larmor frequency
$\nu_\textrm{L}=\gamma_\textrm{N}B_\textrm{z}/(2\pi)$, where
$\gamma_\textrm{N}$ is the nuclear gyromagnetic ratio. Strain
induced quadrupolar effects give rise to small additional shifts
$m^2h\nu_\textrm{Q}/2$ leading to a triplet of dipole-allowed
transitions with splitting $\nu_\textrm{Q}$. The frequencies of
the dipole-allowed transitions are probed by measuring the NMR
spectrum as shown in Fig.~\ref{fig1}c: for a typical individual
dot B1 we find $\nu_\textrm{L}\sim32.94$~MHz and
$\nu_\textrm{Q}\sim65$~kHz for $^{75}$As nuclei at
$B_\textrm{z}\sim4.5$~T. Two features of the nuclear spin spectrum
are important for this work: (i) $\nu_\textrm{Q}\ll\nu_\textrm{L}$
in a wide range of magnetic fields, so that the nuclear spin
levels are nearly equidistant allowing straightforward use of the
nuclear spin temperature concept \cite{GoldmanBook}, (ii) the NMR
triplet is well resolved, providing access to individual spin
transitions and eventually allowing the spin temperature to be
measured.

The collective state of the nuclear spin bath induced via optical
cooling can be characterized by the population probabilities
$p_{m}$ of the levels with nuclear spin projections $m$. In
thermal equilibrium the system is described by the canonical
Boltzmann distribution:
\begin{eqnarray}
p_m = e^{m \beta}/\sum_{m=-I}^{+I}e^{m \beta},\label{eq:NucTemp}
\end{eqnarray}
where dimensionless inverse temperature $\beta=h
\nu_L/k_\textrm{b} T_\textrm{N}$ is introduced and $k_\textrm{b}$
is the Boltzmann constant. For spin $I$=1/2 any statistical
distribution is described by Eq. \ref{eq:NucTemp} with some
$T_\textrm{N}$. However, for $I>$1/2 the nuclear spin temperature
hypothesis of Eq. \ref{eq:NucTemp} is a non-trivial statement
implying existence of equilibration mechanisms which in turn
require sufficiently ''complex'' interactions that can couple all
states of the system leaving the total energy as the only constant
of motion \cite{GoldmanBook}. The polarization degree
\begin{equation}
P_\textrm{N}=\sum_{m=-I}^{+I}mp_{m}/I.\label{eq:PN}
\end{equation}
is uniquely related to $\beta$ and $T_\textrm{N}$ when $p_{m}$ are
given by Eq.~\ref{eq:NucTemp}. An example of a Boltzmann
distribution corresponding to $P_\textrm{N}\approx+80$\% is
sketched in Fig.~\ref{fig1}d for $I$=3/2: most nuclei are in a
$m=+3/2$ state with less than 2\% occupying the $m=-3/2$ state.

Probing $p_{m}$ is achieved by their manipulation with
radio-frequency (rf) pulses as demonstrated in
Figs.~\ref{fig1}e,f. This is possible since the optically-detected
hyperfine shift depends on $P_\textrm{N}$ and hence on $p_{m}$
(Eqs. \ref{eq:Ehf}, \ref{eq:PN}). When, for example, a long rf
pulse with a rectangular spectrum exciting selectively the
$+1/2\leftrightarrow+3/2$ NMR transition [Fig.~\ref{fig1}f top] is
applied, the $p_{+1/2}$ and $p_{+3/2}$ populations are equalized
[Fig.~\ref{fig1}e top] becoming $(p_{+3/2}+p_{+1/2})/2$. The
resulting change in hyperfine shift is $\Delta
E_\textrm{hf}^{+1/2\leftrightarrow+3/2}=-kA
(p_{+3/2}-p_{+1/2})/2$. Similarly, by saturating any
$m\leftrightarrow m+1$ NMR transition it is possible to find the
corresponding $(p_{m+1}-p_{m})$ difference (see Methods).
Simultaneous saturation of two transitions $m\leftrightarrow m+2$
[Figs.~\ref{fig1}e,f middle] reveals the $(p_{m+2}-p_{m})$
differences, while saturation of all three transitions
[Figs.~\ref{fig1}e,f bottom] yields the hyperfine shift variation
$\Delta E_\textrm{hf}^{-3/2\leftrightarrow+3/2}$ proportional to
the polarization degree $P_\textrm{N}$. Thus the statistical
distribution $p_{m}$ can be reconstructed from the optically
detected hyperfine shifts induced by selective manipulation of
$p_{m}$ with rf pulses.

\begin{figure}[h]
\includegraphics[bb=33pt 50pt 422pt 390pt, width=8.5cm]{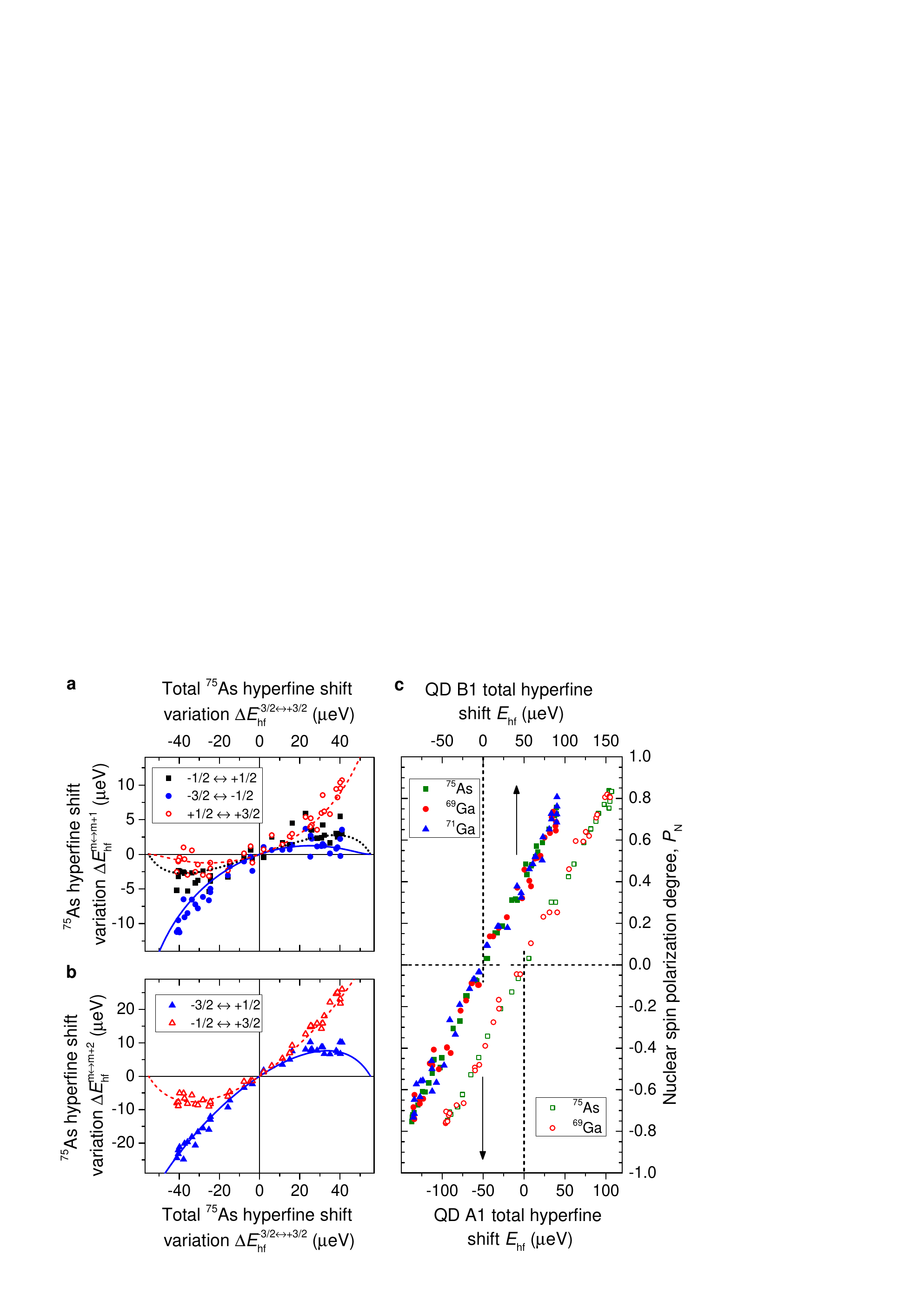}
\caption{\textbf{Probing nuclear spin temperature in a quantum
dot.} \textbf{a,} Symbols show hyperfine shifts induced by
selective saturation of each of the three dipolar NMR transitions
of the spin-3/2 $^{75}$As nuclei as a function of the hyperfine
shift measured by simultaneous saturation of all NMR transitions
in an individual quantum dot B1 at $B_z\approx$4.5~T. \textbf{b,}
Same experiment with simultaneous saturation of two NMR
transitions provides better signal to noise ratio. Lines in
\textbf{a} and \textbf{b} show model calculation with hyperfine
constant as the only fitting parameter. The largest polarization
of $^{75}$As achieved in this experiment
$P_\textrm{N}\approx\pm74\%$ corresponds to nuclear spin
temperature $T_\textrm{N}\approx\pm1.3$~mK. \textbf{c,}
polarization degree $P_\textrm{N}$ of each isotope as a function
of the total hyperfine shift $E_\textrm{hf}$ of all isotopes in QD
B1 (top scale, full symbols) and A1 (bottom scale, open symbols).
All experimental plots in (\textbf{a-c}) are obtained by varying
the degree of circular polarization of the optical pumping.}
\label{fig2}
\end{figure}

Experimental probing of the nuclear spin populations $p_{m}$ is
demonstrated in Fig.~\ref{fig2}a for $^{75}$As nuclei in quantum
dot B1. After optical cooling with light of a variable circular
polarization degree, nuclear spin transitions are saturated
selectively with rf. The resulting changes in hyperfine shifts
$\Delta E_\textrm{hf}^{m\leftrightarrow m+1}$ are shown by the
symbols as a function of the total $^{75}$As hyperfine shift
$\Delta E_\textrm{hf}^{-3/2\leftrightarrow+3/2}$ variation. At
large positive or negative $\Delta
E_\textrm{hf}^{-3/2\leftrightarrow+3/2}$, the $\Delta
E_\textrm{hf}^{+1/2\leftrightarrow+3/2}$ signal is distinctly
larger than $\Delta E_\textrm{hf}^{-3/2\leftrightarrow-1/2}$. Such
difference is even more pronounced in the $\Delta
E_\textrm{hf}^{m\leftrightarrow m+2}$ shifts (Fig.~\ref{fig2}b)
and implies $(p_{+3/2}-p_{+1/2})>(p_{-1/2}-p_{-3/2})$, which is a
clear evidence that large, comparable to unity $|P_\textrm{N}|$ is
induced via optical cooling: indeed, in the high temperature limit
of Boltzmann distribution (large $T_\textrm{N}$ in Eq.
\ref{eq:NucTemp} and small $|P_\textrm{N}|\ll1$), the
probabilities $p_m$ depend linearly on $m$ and all the
$(p_{m+1}-p_{m})$ differences must be equal.

For quantitative analysis we assume the spin temperature
hypothesis (Eq.~\ref{eq:NucTemp}) so that any rf induced hyperfine
shift variation $\Delta E_\textrm{hf}^{m\leftrightarrow m+n}$ is
expressed in terms of $T_\textrm{N}$ and $k A$ (see Methods). By
varying $T_\textrm{N}$ from $-\infty$ to $+\infty$ we obtain
parametric plots of all $\Delta E_\textrm{hf}^{m\leftrightarrow
m+n}$ as a function of $\Delta
E_\textrm{hf}^{-3/2\leftrightarrow+3/2}$ with the $k A$ product
used as the sole fitting parameter. For quantum dot B1 all five
model curves (shown by the lines in Figs.~\ref{fig2}a,b) are in
good agreement with experiment for $k A=36.9\pm1.6~\mu$eV. This
confirms that the nuclear spin state produced by optical cooling
is described by a nuclear spin temperature $T_\textrm{N}$, the
smallest achieved absolute temperature is found to be
$T_\textrm{N}\approx\pm1.3$~mK.

The measurements of Fig.~\ref{fig2}b were repeated on different
isotopes in several quantum dots each time in good agreement with
the spin temperature hypothesis. With $k A$ derived from fitting,
the nuclear spin polarization degree can be obtained as
$P_\textrm{N}=-\Delta E_\textrm{hf}^{-3/2\leftrightarrow+3/2}/(k A
I)$ for each experimental point. Fig.~\ref{fig2}c (top scale)
shows $P_\textrm{N}$ of $^{75}$As, $^{69}$Ga and $^{71}$Ga as a
function of the total hyperfine shift produced by all isotopes in
dot B1. Similar results for another quantum dot (A1) are shown as
well (bottom scale). $P_\textrm{N}$ up to $\sim\pm$80\% is
achieved and to our knowledge is the largest reported for III-V
quantum dots. Two factors are at play here. (i) The efficiency of
nuclear spin cooling in the studied GaAs/AlGaAs nanohole dots is
somewhat higher than in the previous studies: for example, the
total Overhauser shifts $E_\textrm{hf}=\pm100~\mu$eV observed here
and corresponding to $P_\textrm{N}\sim\pm$80\% can be directly
compared to $E_\textrm{hf}=\pm90~\mu$eV observed in GaAs/AlGaAs
fluctuation dots \cite{GammonPRL2001}. (ii) What is more
important, our measurement of $P_\textrm{N}$ does not depend on
the uncertainties in hyperfine constants $A$ and dot structural
parameter $k$ -- it is likely that $P_\textrm{N}$ was
underestimated in all earlier studies.

From Fig.~\ref{fig2}(c) we find that optical cooling produces the
same $P_\textrm{N}$ for all isotopes. Since the Larmor frequencies
$\nu_\textrm{L}$ of the isotopes are significantly different (a
factor of $\sim$1.78 for $^{71}$Ga and $^{75}$As) their spin
temperatures $T_\textrm{N}$ are different too. In other words,
optical cooling leaves the Zeeman reservoir of each isotope in a
state of internal thermal equilibrium, but out of equilibrium with
other isotopes. Detailed measurements conducted on Ga and As
isotopes of several quantum dots in the range $B_z=4.5-8.5$~T
suggest that while $T_\textrm{N}$ varies, the inverse spin
temperature $\beta$ is invariant for the given optical pumping
conditions (with maximum $|\beta|\approx1.43$ corresponding to
$P_\textrm{N}\approx80\%$). This can be understood assuming that
optical cooling of the nuclei arises purely from the hyperfine
induced flip-flop processes where the change in the electron spin
projection $S_z$ by $\pm1$ is accompanied by a $\mp1$ change in
the nuclear spin projection $m$. Furthermore, each nucleus behaves
independently interacting only with the electron spin
\cite{PhysRevB.75.155324,Dyakonov1974}, leading to an invariant
nuclear $p_{m+1}/p_{m}=e^\beta$ determined only by the properties
of the spin polarized electrons. From the thermodynamics
perspective, the isotopic difference in $T_\textrm{N}$ can be seen
to result from the insufficient ''complexity'' of the
electron-nuclear flip-flop process: in particular it does not
provide enough interaction pathways to equilibrate the Zeeman
reservoirs of different isotopes.

The above findings are contrary to the theoretical predictions
that quantum coherence between different nuclei gives rise to
so-called nuclear dark states, limiting the maximum achievable
$|P_\textrm{N}|$ in quantum dots
\cite{ImamogluPRL2003,PhysRevB.75.155324,HildmannPRB2014}. It has
been proposed that small perturbations of the electron
wavefunction must be sufficient to disrupt the dark states and
enable further nuclear spin cooling
\cite{ImamogluPRL2003,PhysRevB.75.155324}. We have performed
experiments where the electron-nuclear spin system is perturbed by
periodically interrupting the optical pumping (which empties the
dot) for periods up to 120~ms that are much longer than the
nuclear spin coherence times $T_2<5$~ms
\cite{ChekhovichNatComm2015} and are thus sufficient for the
nuclei to dephase. However no effect of interruption was found
suggesting that nuclear dark states are not a limiting factor for
achieving $|P_\textrm{N}|$ up to 80\%. We argue that the optical
cooling process already has a natural mechanism disrupting any
nuclear spin coherence, for example the fluctuating electron
hyperfine Knight field, which prevents dark state formation.

With ''dark'' nuclear states ruled out for the studied structures,
it is important to understand what limits $|P_\textrm{N}|$ at
$\approx$80\% and hence find a way to achieve
$|P_\textrm{N}|\approx$100\%. To this end we examine the
dependence of $|P_\textrm{N}|$ on the power and wavelength of the
light used to cool the nuclear spins (see Supplementary Note 3).
The largest $|P_\textrm{N}|$ up to $80$~\% (such as observed in
Fig.~\ref{fig2}) is achieved by exciting $\sim$55~meV above the
exciton ground state. Under such conditions, the limitations to
$|P_\textrm{N}|$ may arise from the loss of electron spin
polarization during energy relaxation, and competing effects of
heavy and light hole excitation \cite{PhysRevB.90.125301}.
Moreover the maximum $|P_\textrm{N}|$ is observed under optical
excitation powers $\sim$1000 times larger than the saturation
power of the quantum dot photoluminescence, i.e. spin cooling is
related to multiexciton complexes rather than the ground state
exciton. Under such conditions optically induced nuclear spin
relaxation \cite{PhysRevB.77.245201} could be a significant
limiting factor. By contrast nuclear spin relaxation in the
absence of optical excitation is found to be slow ($T_1>$500~s)
\cite{Ulhaq} and can be ruled out as a limitation to
$|P_\textrm{N}|$. Further experimental and theoretical studies are
needed to disentangle the role of the various effects, in
particular explaining why nuclear spin cooling turns out to be
most efficient at surprisingly high optical powers and excess
photon energies.

We finally turn to derivation of the GaAs hyperfine constants,
which are defined as \cite{Gueron,Paget1977}
\begin{equation}
A =
(2\mu_0/3)\hbar\gamma_\textrm{N}g_\textrm{e}\mu_\textrm{b}|\psi(0)|^2,\label{eq:Adef}
\end{equation}
where $g_\textrm{e}\approx2$ is the free electron g-factor,
$\mu_0$ is magnetic constant, $\mu_\textrm{b}$ is Bohr magneton,
$\hbar=h/(2\pi)$, and the electron wavefunction density
$|\psi(0)|^2$ at the nucleus of the corresponding isotope is the
only parameter characterizing GaAs as a material. By definition,
when nuclei with spin $I$ are 100\% abundant and fully polarized
($|P_\textrm{N}|=1$) the electron hyperfine shift is
$E_\textrm{hf}=AI$ regardless of the electron envelope
wavefunction \cite{Paget1977}. However, the effective hyperfine
constant $k A$ obtained from experiment is reduced by a factor
$k\leq1$ which is a product of the isotope natural abundance
$\rho\leq1$, molar fraction $x\leq1$ and a further factor $\leq1$
describing the non-uniformities of the nuclear polarization and
electron envelope wavefunction.

In order to estimate $k$ we study dots with different localization
energy. We find that $k A$ of the dots with deeper confining
potential (emitting at $\sim$1.58~eV) is larger by a factor
$\sim$1.09 than for the dots with weaker confinement (emitting at
$\sim$1.63~eV). Such a difference can be explained if we assume
that optical cooling produces uniform nuclear polarization
$|P_\textrm{N}|$ within the GaAs quantum dot volume while the
nuclei of the AlGaAs barrier remain nearly unpolarized
($|P_\textrm{N}|\approx0$) due to the quadrupolar induced
suppression of spin diffusion at the GaAs/AlGaAs interface
\cite{Ulhaq}. In such case $k=x\rho W$, where $W$ is the portion
of the electron density within GaAs. Solving the Schrodinger's
equation we find $W\approx0.94$ for dots emitting at $\sim$1.58~eV
and $W\approx0.88$ for dots emitting at $\sim$1.63~eV (see
Supplementary Note 4). The ratio of the calculated $W$ values is
$\approx$1.07 in agreement with the $\approx$1.09 ratio of the
measured $k A$, confirming the assumption that
$P_\textrm{N}\approx0$ outside the dot and $P_\textrm{N}\neq0$
within the dot. Using these $k$ values we correct $A$ and
$|\psi(0)|^2$ for each studied dot and after averaging we finally
obtain the 95\%-confidence estimates
$|\psi_\textrm{As}(0)|^2$=(9.25$\pm$0.20)$\times$10$^{31}$~m$^{-3}$
and
$|\psi_\textrm{Ga}(0)|^2$=(6.57$\pm$0.25)$\times$10$^{31}$~m$^{-3}$.
The corresponding hyperfine constants are $A$=43.5$\pm$0.9~$\mu$eV
($^{75}$As), $A$=43.1$\pm$1.6~$\mu$eV ($^{69}$Ga), and
$A$=54.8$\pm$2.1~$\mu$eV ($^{71}$Ga). Our experimental values for
$|\psi(0)|^2$ generally agree with the original estimates of Paget
et al. \cite{Paget1977} based on the studies on InSb
\cite{Gueron}. On the other hand the ratio
$|\psi_\textrm{As}(0)|^2/|\psi_\textrm{Ga}(0)|^2\approx1.41$ which
has the highest accuracy as it does not depended on assumptions
about $k$ is somewhat smaller than 1.69, based on the estimates in
Ref.~\cite{Paget1977}.

In conclusion, we have demonstrated a direct measurement of the
temperature of the nuclear spin Zeeman reservoir in weakly
strained quantum dots where individual NMR transitions are well
resolved. While NMR transitions can not be fully resolved in
self-assembled dots due to the strain inhomogeneity
\cite{ChekhovichNatNano2012}, the techniques developed here can be
applied to the weakly broadened $m=\pm1/2$ subensemble providing a
way to explore mesoscopic nuclear spin thermodynamics both in high
magnetic fields as studied here, and in low fields where the spin
temperature concept does not apply to the full spin 3/2 or 9/2
manifolds \cite{Maletinsky2009}. Nuclear polarizations up to 80\%
are achieved and are still not limited by the nuclear coherence
effects. We expect that techniques for monitoring the state of the
nuclear spin bath reported here will stimulate further effort
towards initializing the mesoscopic nuclear spin bath in a highly
polarized state. Both experimental and theoretical work would be
needed to understand the detailed role of various factors limiting
the nuclear polarization as well as engineering quantum dot
structures and protocols allowing for maximal nuclear spin cooling
efficiency.

\section{Methods}

\textbf{Experimental techniques.} We use neutral quantum dots,
i.e. without optical excitation the dots are empty. In all
measurements we use the \textit{Optical cooling - rf
depolarization - Optical readout} protocol described previously in
detail \cite{ChekhovichNatNano2012,ChekhovichNatComm2015}. The rf
depolarization is performed in the absence of optical excitation.
The role of the short optical probe pulse is to excite
photoluminescence whose spectrum is then analyzed with a double
spectrometer and a CCD camera.

Photoluminescence of a neutral QD results from recombination of an
electron with spin up ($\uparrow$) or down ($\downarrow$) and a
hole with spin up ($\Uparrow$) or down ($\Downarrow$) along the
sample growth direction ($z$ axis). We observe emission of both
the ''bright'' excitons $\ket{\Uparrow\downarrow}$,
$\ket{\Downarrow\uparrow}$ and ''dark'' excitons
$\ket{\Uparrow\uparrow}$, $\ket{\Downarrow\downarrow}$ that gain
oscillator strength from the bright states due to the reduced
quantum dot symmetry.

The net nuclear spin polarization shifts the energies of the
exciton states. The shifts are dominated by the sign of the
electron spin $z$ projection, they are $\approx+E_\textrm{hf}/2$
for $\ket{\Downarrow\uparrow}$ and $\ket{\Uparrow\uparrow}$
states, and $\approx-E_\textrm{hf}/2$ for
$\ket{\Uparrow\downarrow}$ and $\ket{\Downarrow\downarrow}$
states. In order to determine $E_\textrm{hf}$ accurately we
measure the energy difference of the $\ket{\Downarrow\uparrow}$
and $\ket{\Downarrow\downarrow}$ bright and dark states (or of the
$\ket{\Uparrow\uparrow}$ and $\ket{\Uparrow\downarrow}$ dark and
bright states). In this way we eliminate any contribution of the
hole hyperfine interaction, as well as any simultaneous shifts of
all exciton states arising e.g. from charge fluctuations in the
dot vicinity.

\textbf{Relation between the nuclear spin population probabilities
$p_m$ and the optically detected hyperfine shifts.} In experiments
we use long and weak (no Rabi oscillations) radio frequency (rf)
excitation. If rf is resonant with the NMR transition between
states $m$ and $m+1$ its effect is to change and equalize the
populations of these states so that $p_{m}, p_{m+1}\rightarrow
(p_{m}+p_{m+1})/2$, while population probabilities of all other
nuclear spin states remain unchanged. One can then use Eqs
\ref{eq:Ehf} and \ref{eq:PN} to calculated the change in the
optically detected hyperfine shift $\Delta E_\textrm{hf}$ arising
from such manipulation of $p_{m}$. For example for $m=+1/2$ and
$m+1=+3/2$ we calculate as follows: $\Delta
E_\textrm{hf}^{+1/2\leftrightarrow+3/2}=kA\left[(+\frac{3}{2})\frac{p_{+3/2}+p_{+1/2}}{2}+(+\frac{1}{2})\frac{p_{+3/2}+p_{+1/2}}{2}\right]-kA\left[(+\frac{3}{2})p_{+3/2}+(+\frac{1}{2})p_{+1/2}\right]=-kA
(p_{+3/2}-p_{+1/2})/2$, i.e. the hyperfine shift $\Delta
E_\textrm{hf}$ depends only on the difference in the initial
populations of the states excited with rf.

In a similar way, simultaneous saturation of the NMR transitions
$m \leftrightarrow m+1$ and $m+1 \leftrightarrow m+2$ leads to the
following redistribution of the populations: $p_{m}, p_{m+1},
p_{m+2}\rightarrow (p_{m}+p_{m+1}+p_{m+2})/3$. Saturation of all 3
NMR transition of spin $I$=3/2 nuclei leads to $p_{-3/2},
p_{-1/2}, p_{+1/2}, p_{+3/2}\rightarrow
(p_{-3/2}+p_{-1/2}+p_{+1/2}+p_{+3/2})/4 = 1/4$, where the last
equality is due to normalization $\sum_{m=-I}^{I}p_m=1$. Using
Eqs. \ref{eq:Ehf}, \ref{eq:PN} we evaluate the changes in the
hyperfine shift $\Delta E_\textrm{hf}$ for each case:
\begin{eqnarray}
\begin{aligned}
&\Delta E_\textrm{hf}^{m\leftrightarrow m+1}=-k A
(p_{m+1}-p_{m})/2=\\
&\quad\quad\quad=-k A\frac{e^{(m+1)\beta}-e^{m\beta}}{4\cosh(\beta/2)+4\cosh(3\beta/2)},\\
&\Delta E_\textrm{hf}^{m\leftrightarrow m+2}=-k A (p_{m+2}-p_{m})=\\
&\quad\quad\quad=-k A e^{(m+1)\beta} \sinh(\beta/2)/\cosh(\beta),\\
&\Delta E_\textrm{hf}^{-I\leftrightarrow +I}=-k A P_\textrm{N}I=\\
&\quad\quad\quad=-k
A[3/2+1/\cosh(\beta)]\tanh(\beta/2).\label{eq:DEhf}
\end{aligned}
\end{eqnarray}
The last expression in each of these equations is obtained by
substituting the Boltzmann distribution (Eq.~\ref{eq:NucTemp}).

Unlike $\Delta E_\textrm{hf}^{m\leftrightarrow m+1}$ and $\Delta
E_\textrm{hf}^{m\leftrightarrow m+2}$, the $\Delta
E_\textrm{hf}^{-I\leftrightarrow +I}$ variation is a monotonic
function of $\beta$ with $\beta\in (-\infty,+\infty)$. It is thus
possible to express $\beta$ in terms of $\Delta
E_\textrm{hf}^{-I\leftrightarrow +I}$ and then substitute it into
expressions for $\Delta E_\textrm{hf}^{m\leftrightarrow m+1}$ and
$\Delta E_\textrm{hf}^{m\leftrightarrow m+2}$, eliminating
$\beta$. Since there is no analytical solution we do this by
making parametric plots of $\Delta E_\textrm{hf}^{m\leftrightarrow
m+1}$ and $\Delta E_\textrm{hf}^{m\leftrightarrow m+2}$ as a
function of $\Delta E_\textrm{hf}^{-I\leftrightarrow +I}$ such as
shown in Figs. \ref{fig2}a,b by the lines. The $k A$ product is an
overall scaling factor used as a parameter to fit the experimental
data. One can see that such fitting can be achieved reliably only
because large $P_\textrm{N}$ is reached in experiment: only then
there are pronounced asymmetries and nonlinearities in the $\Delta
E_\textrm{hf}^{m\leftrightarrow m+1}(\Delta
E_\textrm{hf}^{-I\leftrightarrow +I})$ and $\Delta
E_\textrm{hf}^{m\leftrightarrow m+2}(\Delta
E_\textrm{hf}^{-I\leftrightarrow +I})$ dependencies (observed in
Figs. \ref{fig2}a,b). By contrast, in the high temperature limit
($\beta \rightarrow 0$, $P_\textrm{N} \rightarrow 0$)
Eq.~\ref{eq:DEhf} yields linear dependencies $\Delta
E_\textrm{hf}^{m\leftrightarrow m+1}=(1/10)\Delta
E_\textrm{hf}^{-I\leftrightarrow +I}$, $\Delta
E_\textrm{hf}^{m\leftrightarrow m+2}=(4/10)\Delta
E_\textrm{hf}^{-I\leftrightarrow +I}$ independent of $m$, so that
experiment can be described with any $k A$, making fitting
impossible.

Once $k A$ is obtained from fitting, one can use the last of
Eq.~\ref{eq:DEhf} to uniquely relate the experimentally measured
$\Delta E_\textrm{hf}^{-I\leftrightarrow +I}$, and the quantities
of interest such as polarization degree $P_\textrm{N}$, the
inverse temperature $\beta$, and the nuclear spin temperature
$T_\textrm{N}$ itself.

\textbf{ACKNOWLEDGMENTS} The authors are grateful to Armando
Rastelli (Linz), Yongheng Huo (Hefei) and Andreas Waeber (TU
Munich) for fruitful discussions. This work has been supported by
the EPSRC Programme Grant EP/J007544/1. E.A.C. was supported by a
University of Sheffield Vice-Chancellor's Fellowship and a Royal
Society University Research Fellowship.



\renewcommand{\thesection}{Supplementary Note \arabic{section}}
\setcounter{section}{0}
\renewcommand{\thefigure}{\arabic{figure}}
\renewcommand{\figurename}{Supplementary Figure}
\setcounter{figure}{0}
\renewcommand{\theequation}{\arabic{equation}}
\setcounter{equation}{0}
\renewcommand{\thetable}{Supplementary Table \arabic{table}}
\setcounter{table}{0}

\renewcommand{\citenumfont}[1]{S#1}
\makeatletter
\renewcommand{\@biblabel}[1]{S#1.}
\makeatother

\pagebreak \pagenumbering{arabic}

\section*{SUPPLEMENTARY INFORMATION}

\section{Sample structure.}

The sample growth process has been reported in detail
previously\cite{Atkinson2012}. Here we give a brief description.

The nanohole filled droplet epitaxial quantum dot sample is grown
via solid molecular beam epitaxy (MBE). The GaAs buffer layer is
first grown and is followed by deposition of 11 monolayers of Ga
at $520~^\circ$C forming Ga droplets. The droplets are then
annealed under As flux resulting in crystallization and eventual
formation of nanoholes due to As dissolution and Ga diffusion. The
holes are then filled by depositing $7~$nm of
Al$_{0.44}$Ga$_{0.56}$As forming the bottom barrier. This is
followed by deposition of $3.5~$nm of GaAs. Due to the difference
in migration rates of Ga and Al, GaAs redistributes towards the
bottom of the nanohole. Such ''in-filling'' of the nanoholes
results in formation of inverted QDs. Finally the dots are capped
by a $112$~nm Al$_{0.33}$Ga$_{0.67}$As top barrier and $20~$nm
GaAs layer.

\section{Optical detection of the hyperfine shifts.\label{SI:EhfDetection}}

In all nuclear spin cooling experiments we use \textit{Optical
pump - rf depolarization - Optical probe} protocol which has been
described in detail previously
\cite{ChekhovichNatNano2012,ChekhovichNatPhy2013,MunschNMR,ChekhovichNatComm2015,ChekhovichNatPhy2016}.

Optical pumping is implemented using Ti:Sap laser emission of a
variable wavelength and power. The largest nuclear spin
polarization $|P_\textrm{N}|\approx80$~\% is achieved at
excitation power of $\sim3000$~$\mu$W and photon energy of
$\sim$1.63~eV for quantum dots emitting at $\sim$1.58~eV and at a
photon energy of $\sim$1.645~eV for quantum dots emitting at
$\sim$1.63~eV (see further discussion in \ref{SI:DNPVSLambda}).
The duration of the optical pump is typically
$t_\textrm{Pump}=10-12$~s, sufficiently long to induce a steady
state nuclear spin polarization. A combination of a half-wave and
a quarter-wave plates is used to control the degree of circular
polarization of the pump laser from -1 to +1 (polarization is
varied from $\sigma^-$ to $\sigma^+$ through intermediate
elliptical polarizations): in this way the nuclear spin
polarization degree could be controlled gradually between its
maximum negative and positive values as shown in Fig. 2c of the
main text.

Selective depolarization of the nuclei is achieved by applying
radio frequency oscillating magnetic field with rectangular
spectral profile as shown in Fig.~1f of the main text. Such
rectangular bands are constructed from frequency combs with comb
spacing $f_\textrm{MS}=125$~Hz much smaller than the homogeneous
NMR linewidths, so that the combs are equivalent to white noise
\cite{ChekhovichNatPhy2016}. Typical depolarization time is
$t_\textrm{rf}=1$~s.

After rf depolarization, a HeNe (632.8~nm) probe laser pulse is
used to excite quantum dot photoluminescence (PL). The duration of
the probe is typically $t_\textrm{Probe}=40-100$~ms and is chosen
short enough to minimize its effect on the nuclear spin
polarization. Exemplary probe PL spectra are shown in
Supplementary Fig. \ref{Fig:SuppPLSpec}.

\begin{figure}[h]
\includegraphics[bb=23pt 23pt 556pt 655pt, width=9cm]{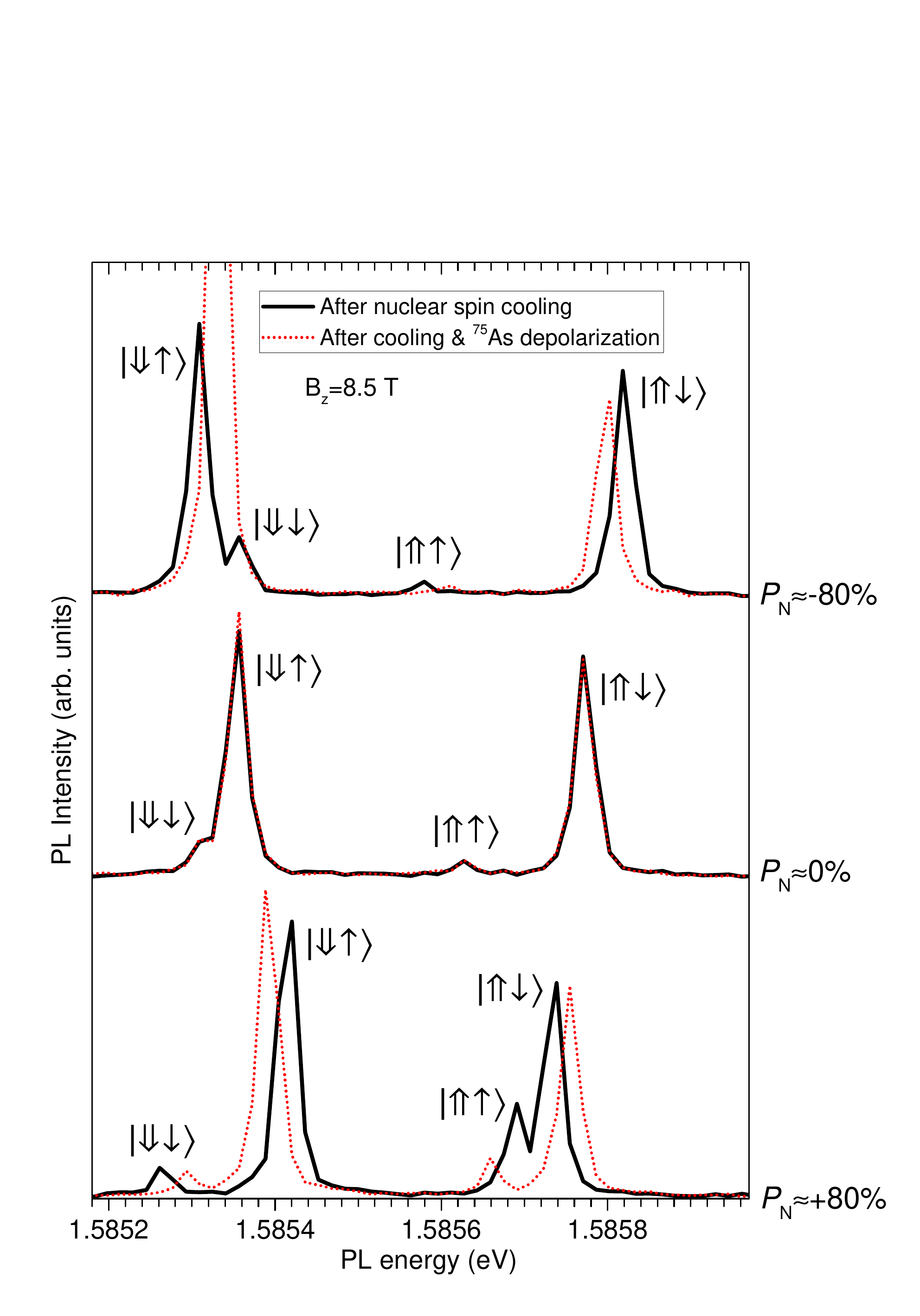}
\caption{\textbf{Optical detection of the hyperfine shifts.}
Photoluminescence (PL) spectra of a single neutral GaAs/AlGaAs
nanohole filled quantum dot in magnetic field $B_z\approx8.5$~T
recorded after cooling the quantum dot nuclear spins to different
polarization degrees $P_\textrm{N}$ (solid lines). The dashed
lines show PL spectra recorded after cooing followed by selective
depolarization of the $^{75}$As nuclei. In each spectrum four PL
lines are observed corresponding to all possible combinations of
the electron spin states ($\uparrow$, $\downarrow$) and hole
states ($\Uparrow$, $\Downarrow$) forming two bright excitons
$\left|\Uparrow\downarrow\right>$,
$\left|\Downarrow\uparrow\right>$ and two dark excitons
$\left|\Uparrow\uparrow\right>$,
$\left|\Downarrow\downarrow\right>$ that have finite admixture of
bright states making dark states visible in PL
\cite{Bayer2000,ChekhovichPRL2011,Puebla2013}. The splitting of
the
$\left|\Downarrow\uparrow\right>$-$\left|\Downarrow\downarrow\right>$
exciton pair (or the
$\left|\Uparrow\uparrow\right>$-$\left|\Uparrow\downarrow\right>$
pair) depends on the nuclear spin polarization. The change in this
splitting induced by rf depolarization yields the electron
hyperfine shift $E_\textrm{hf}$ of the corresponding isotope
($^{75}$As in this case). At certain levels of nuclear
polarization dark and bright states overlap and anticross (e.g.
$\left|\Downarrow\uparrow\right>$-$\left|\Downarrow\downarrow\right>$
at large negative $P_\textrm{N}$ or
$\left|\Uparrow\uparrow\right>$-$\left|\Uparrow\downarrow\right>$
at large positive $P_\textrm{N}$), in such cases the other
dark-bright pair can still be used to measure $E_\textrm{hf}$.}
\label{Fig:SuppPLSpec}
\end{figure}

For the neutral dots studied in this work in presence of large
magnetic field along the growth axis, the PL spectra exhibit
emission of all four excitonic states formed by an electron with
spin up or down ($\uparrow$, $\downarrow$) and a hole with spin up
or down ($\Uparrow$, $\Downarrow$): there are two bright excitons
$\left|\Uparrow\downarrow\right>$,
$\left|\Downarrow\uparrow\right>$ and two dark excitons
$\left|\Uparrow\uparrow\right>$,
$\left|\Downarrow\downarrow\right>$ that have finite admixture
from bright states making them visible in PL \cite{Bayer2000}. Due
to the hyperfine interaction, nuclear spin polarization shifts the
energy of each exciton state according to its electron spin and
hole spin directions. The contribution of the hole hyperfine
interaction is small but not negligible
\cite{ChekhovichNatPhy2013}. In order to exclude it and measure
pure electron hyperfine shifts, we use the splitting in energies
of a bright and a dark exciton with the same hole spin projection,
for example a
$\left|\Uparrow\uparrow\right>$-$\left|\Uparrow\downarrow\right>$
pair of states. Such splitting equals the total electron hyperfine
shift $E_\textrm{hf}$ plus a constant Zeeman splitting determined
by the electron and hole $g$-factors. In order to eliminate the
Zeeman contribution and obtain the absolute value of the hyperfine
shift for a selected NMR transition of a selected isotope we
perform a differential measurement: The probe spectra are measured
with rf depolarization (dashed lines in Supplementary
Fig.~\ref{Fig:SuppPLSpec}) and without (solid lines in
Supplementary Fig.~\ref{Fig:SuppPLSpec}), the difference in the
dark-bright splitting of the two spectra gives the required
hyperfine shift. For example the total hyperfine shift
$E_\textrm{hf}$ of an isotope is found as $E_\textrm{hf}=-\Delta
E_\textrm{hf}^{-I\leftrightarrow +I}$, where $\Delta
E_\textrm{hf}^{-I\leftrightarrow +I}$ is the change in dark-bright
exciton splitting induced by saturating all NMR transitions (see
main text and Methods).

\section{Dependence of the nuclear spin cooling efficiency on the power and wavelength of the optical pumping.\label{SI:DNPVSLambda}}

\begin{figure}[h]
\includegraphics[bb=23pt 23pt 556pt 400pt, width=12cm]{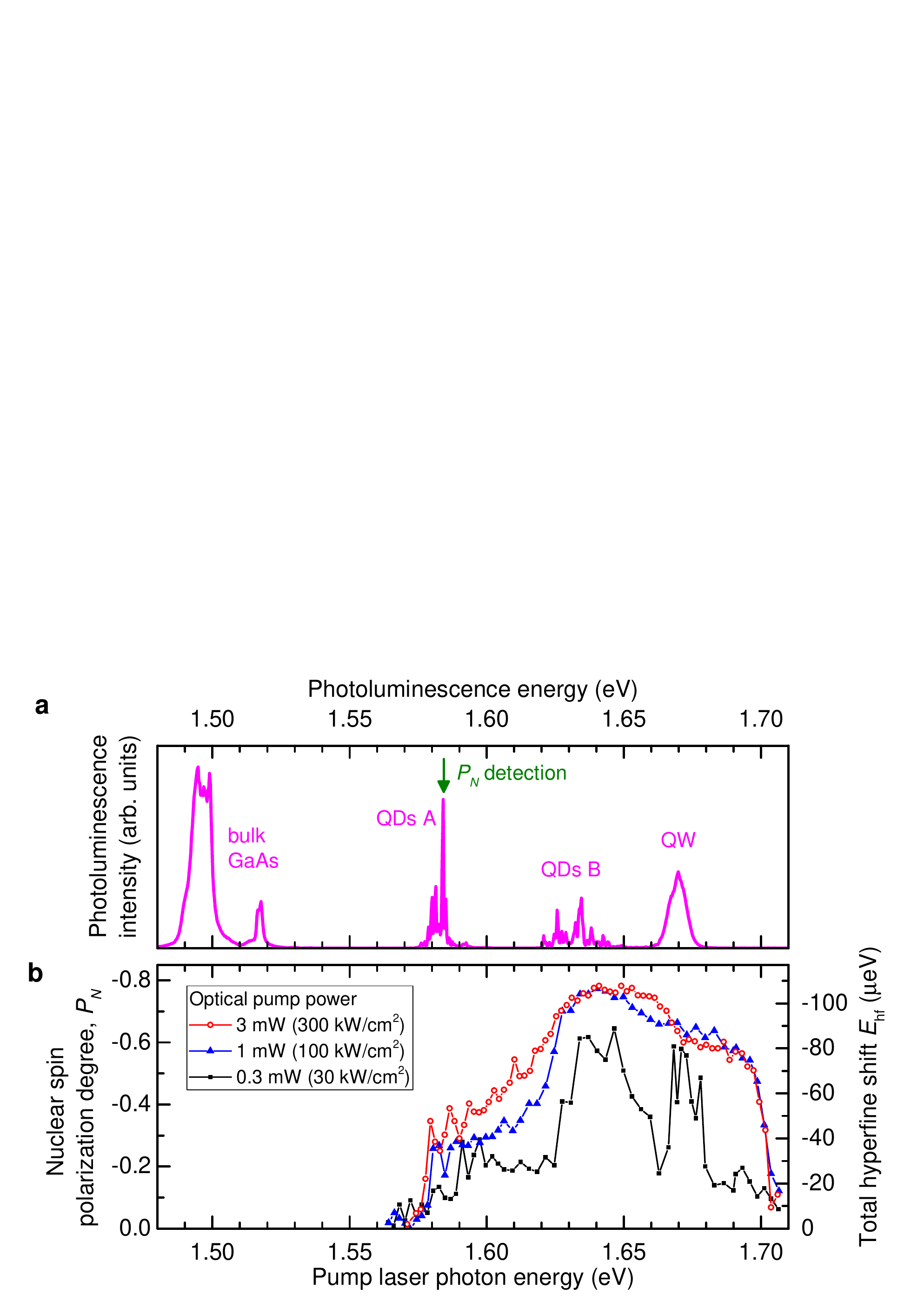}
\caption{\textbf{Dependence of the nuclear spin cooling efficiency
on the power and wavelength of the optical pumping.} \textbf{a,}
Photoluminescence (PL) spectrum measured at $B_z=5~$T and HeNe
laser excitation (632.8~nm). Emission from GaAs substrate,
long-wavelength quantum dots (type A), short-wavelength quantum
dots (type B) and quantum well (QW) are observed. \textbf{b,}
Total optically induced hyperfine shift $E_\textrm{hf}$ and the
corresponding nuclear spin polarization degree $P_\textrm{N}$
measured on one of the dots type A (marked by an arrow) as a
function of optical pumping photon energy at different optical
powers of a circularly polarized laser excitation. The optical
power density is calculated assuming the laser spot area of
1~$\mu$m$^2$.} \label{Fig:FigSuppDNPPLE}
\end{figure}

Supplementary Figure~\ref{Fig:FigSuppDNPPLE}a shows a broad-range
photoluminescence (PL) spectrum of a studied sample measured under
non-resonant laser excitation (at 632.8 nm). Several emission
features are observed and ascribed to (from left to right) bulk
GaAs substrate, long-wavelength quantum dots (type A),
short-wavelength quantum dots (type B) and a wetting layer quantum
well (QW). Supplementary Figure~\ref{Fig:FigSuppDNPPLE}b shows the
total optically induced hyperfine shift $E_\textrm{hf}$ and the
corresponding nuclear spin polarization degree $P_\textrm{N}$
detected on one of the dots type A (marked by an arrow in
Supplementary Figure~\ref{Fig:FigSuppDNPPLE}a) as a function of a
photon energy of a circularly polarized laser at three different
excitation powers.

For the lowest used power of 0.3~mW (corresponding to the surface
power density of $\sim$30~kW/cm$^2$) the result is similar to what
was observed previously in the same structure \cite{Ulhaq} at a
comparable excitation power of 0.5~mW: two broad peaks in nuclear
spin polarization degree $P_\textrm{N}$ detected in a type A dot
are attributed to the resonant optical pumping of the type B dots
($\sim$1.645~eV peak), and the quantum well ($\sim$1.675~eV peak
consisting of sharp features). When the power is increased up to
3~mW, nuclear spin cooling becomes more efficient for the entire
range of the laser photon energies and the broad peaks observed at
0.3~mW broaden further and smear out completely. The largest
$|P_\textrm{N}|$ is observed for a range of energies approximately
corresponding to optical excitation of the type B dots. We thus
argue that in the studied structures, the most efficient nuclear
spin pumping mechanism is via resonant optical injection of spin
polarized excitons into the high-energy type B dots with a
subsequent tunneling and relaxation into the low-energy type A
dots. Measurements at the laser energy of $\sim$1.63~eV (which is
optimal at 3~mW power) for even higher optical powers up to 15~mW
(not shown here) have revealed reduction in $|P_\textrm{N}|$, most
likely arising from local sample heating. It also follows from
Supplementary Figure~\ref{Fig:FigSuppDNPPLE} that for high power
optical pumping, a significant nuclear spin cooling can be induced
for the entire range of energies between QDs type A and type B,
i.e. resonant excitation of type B dots or QW are not the only
mechanisms. We further note that $|P_\textrm{N}|$ up to 30\% is
observed when pumping with energies as low as $\sim$5~meV
\textit{below} the ground state neutral exciton energy of the
studied dot. This can be due to the nuclear spin cooling via
optical pumping of bi-exciton, multi-exciton or (multi-)charged
states. These observations suggest that optical nuclear spin
cooling is a complex process, driven by a combination of various
nuclear spin pumping and nuclear spin depolarization mechanisms.

It is thus evident that a significant further effort is required
in order to understand the mechanisms and engineer the approaches
for achieving even deeper cooling of the nuclear spins with
$|P_\textrm{N}|>80\%$. In this respect, we note that the data
presented in Supplementary Figure~\ref{Fig:FigSuppDNPPLE}b
provides only a snapshot of the nuclear spin cooling phenomena in
the studied dots. Indeed the measurements were conducted with a
relatively broad laser excitation ($\sim$40~GHz) and coarse steps
in photon energy, so that the nuclear spin cooling mechanisms via
resonant \cite{Latta2009,ChekhovichPRL2010} and quasi-resonant
($p$-shell) \cite{Lai2006} optical excitation are yet to be
explored. Since each point in Supplementary
Figure~\ref{Fig:FigSuppDNPPLE}b requires several minutes of PL
spectrum integration, a detailed high resolution exploration of
$|P_\textrm{N}|$ as a function of laser power and photon energy
(or even more detailed measurements with two or more single-mode
tuneable lasers) would require significant experimental effort and
is a subject of further work.

\section{Derivation of electron hyperfine constants $A$: detailed analysis.}
The experimentally measured hyperfine shift $E_\textrm{hf}$
induced by the polarized nuclear spins is defined as the change in
the energy splitting of the $S_\textrm{z}=\pm$1/2 electron spin
levels. The total hyperfine shift is a sum of the hyperfine shifts
induced by different isotopes $i$:
\begin{equation}
E_\textrm{hf} = \sum_i E_\textrm{hf}^{i},\label{eq:EhfTot}
\end{equation}
The hyperfine shift of the $i$-th isotope in a quantum dot is
given by
\begin{equation}
E_\textrm{hf}^{i} = A^i I^i \sum_j \rho^i x^i(\mathbf{r}_j)
|F(\mathbf{r}_j)|^2
P_\textrm{N}(\beta^i(\mathbf{r}_j)),\label{eq:Ehfi}
\end{equation}
where the summation goes over all cationic or anionic (depending
on the type of the isotope $i$) sites $j$ with coordinates
$\mathbf{r}_j$, $A^i$ is the electron hyperfine constant
determined only by the fundamental constants and the density of
the electron Bloch wavefunction $|\psi(0)|^2$ at the nucleus,
$I^i$ is the nuclear spin and $\rho^i$ is the natural abundance of
the $i$-th isotope. The nuclear spin polarization degree
$P_\textrm{N}$ is always defined and is uniquely related (via
Brillouin function) to the dimensionless inverse nuclear spin
temperature $\beta$ if the spin temperature $T_\textrm{N}$ exists
($\beta=h \nu_L/k_\textrm{b} T_\textrm{N}$, where $\nu_L$ is the
nuclear Larmor frequency, and for spin $I$=3/2 the
$P_\textrm{N}(\beta^i)$ is given by the last of Eqs. 5 of the main
text when divided by $-kAI$). The inverse temperature
$\beta^{i}(\mathbf{r}_j)$ and the mole fraction
$x^i(\mathbf{r}_j)$ are not constant in general and depend on
$\mathbf{r}_j$. ($x^i(\mathbf{r}_j)$ is defined as the probability
that the $j$-th site is occupied by an atom of the element to
which the isotope $i$ belongs, e.g. $x$=0.5 for Al and Ga in a
uniform Al$_{0.5}$Ga$_{0.5}$As alloy). $F(\mathbf{r}_j)$ is the
envelope wavefunction of a localized electron state normalized in
a way that $\sum_j |F(\mathbf{r}_j)|^2=1$ when summed over all
cationic or anionic sites of the crystal.

Supplementary Eq. \ref{eq:Ehfi} gives a complete description of
the experimentally observable hyperfine shifts and is the key
expression used in the subsequent analysis. If nuclear spin
temperature exists Supplementary Eq. \ref{eq:Ehfi} can be readily
adjusted to yield the experimentally measurable changes in
hyperfine shifts $\Delta E_\textrm{hf}^{m\leftrightarrow m+1}$ and
$\Delta E_\textrm{hf}^{m\leftrightarrow m+2}$ induced by selective
rf saturation of the NMR transitions, for this
$P_\textrm{N}(\beta)$ needs to be substituted by the corresponding
hyperbolic function of $\beta$ (obtained from Eqs.~5 of the main
text by dividing by $-kAI$).

If $\beta^i$ (and hence $P_\textrm{N}^{i}$) is constant over the
volume of a quantum dot and its vicinity, Supplementary
Eq.~\ref{eq:Ehfi} simplifies to Eq.~1 of the main text
[$E_\textrm{hf}^{i} = k A^{i} I^{i} P_\textrm{N}^{i}$ with $k$
determined only by the structural parameters of the quantum dot
$\rho^i$, $x^i(\mathbf{r}_j)$ and $F(\mathbf{r}_j)$] also leading
to Eqs.~5 of the main text. For the spatially inhomogeneous
$\beta^i$, Eq.~1 of the main text still describes the
experimentally measured hyperfine shifts $E_\textrm{hf}^{i}$ if
$P_\textrm{N}^{i}$ is treated as an average polarization degree,
with $k$ ($0\leq k\leq1$) depending not only on the
$x^i(\mathbf{r}_j)$ and $F(\mathbf{r}_j)$ functions, but also on
the particular form of $\beta^{i}(\mathbf{r}_j)$. In a similar
way, the experimentally measured $\Delta
E_\textrm{hf}^{m\leftrightarrow m+1}$ and $\Delta
E_\textrm{hf}^{m\leftrightarrow m+2}$ can be treated as a result
of averaging over the dot volume. The crucial difference is that
the $\Delta E_\textrm{hf}^{m\leftrightarrow m+1}(\Delta
E_\textrm{hf}^{-I\leftrightarrow +I})$ and $\Delta
E_\textrm{hf}^{m\leftrightarrow m+2}(\Delta
E_\textrm{hf}^{-I\leftrightarrow +I})$ dependencies are not
necessarily described by Eqs. 5 of the main text if the hyperfine
shifts are obtained by averaging over a distribution of
$\beta^{i}(\mathbf{r}_j)$. Since model fitting of the $\Delta
E_\textrm{hf}^{m\leftrightarrow m+1}(\Delta
E_\textrm{hf}^{-I\leftrightarrow +I})$ and $\Delta
E_\textrm{hf}^{m\leftrightarrow m+2}(\Delta
E_\textrm{hf}^{-I\leftrightarrow +I})$ experimental data is
required for the derivation of the nuclear spin temperatures and
hyperfine constants, additional justification of the analysis is
needed and is presented below.

One can see from Supplementary Eq.~\ref{eq:Ehfi} what the
difficulty is: the experimentally measured hyperfine shifts
$E_\textrm{hf}^i$ [left side of the equation] are scalars and can
not on their own give full information on the three-dimensional
distributions $\beta^{i}(\mathbf{r}_j)$, $x^i(\mathbf{r}_j)$ and
$F(\mathbf{r}_j)$ [right side of the equeation], and thus
additional information and/or assumptions about these functions
are needed. The detailed analysis is presented in the subsequent
subsections and can be outlined as follows: (A) the molar
fractions $x^i(\mathbf{r}_j)$ are estimated from the structural
studies on GaAs nanohole quantum dots, (B) the electron envelope
wavefunctions $F(\mathbf{r}_j)$ are calculated numerically by
solving the Schrodinger equation, (C) we show that a particular
form of $\beta^{i}(\mathbf{r}_j)$ is not important and the
derivation of the nuclear spin polarization degrees is robust for
a wide range of distributions of $\beta^{i}$.

\subsection{Effect of the quantum dot structure $x^i(\mathbf{r}_j)$.}

Since arsenic is the only anion in the studied GaAs/AlGaAs
structures its molar fraction is $x^\textrm{As}=1$, simplifying
Eq.~\ref{eq:Ehfi}. As to cations (gallium and aluminium), the
earlier TEM studies on similar sample structures have shown sharp
interfaces between GaAs and AlGaAs layers \cite{Pfeiffer2014}. We
thus use the known molar fractions of aluminium in the barriers to
model $x^i(\mathbf{r}_j)$ as a piece-wise function. Such
approximation is further justified \textit{a posteriori} by the
smallness ($<12\%$) of the fraction of the electron wavefunction
in the AlGaAs layers as confirmed by wavefunction calculations 
(see below in B). Under such conditions, the average cationic
molar fractions probed by the electron are effectively
$x^\textrm{Ga}\approx1$, $x^\textrm{Al}\approx0$ and the
particulars of Al/Ga intermixing profile have little effect on the
hyperfine shifts and the derivation of the nuclear spin
temperatures.

\subsection{Calculation of the electron envelope wavefunction $F(\mathbf{r}_j)$.}

The electron envelope wavefunction $F(\mathbf{r}_j)$ that appears
as a weighting function in Eq.~\ref{eq:Ehfi} is calculated by
solving the Schrodinger equation using effective mass
approximation. We generally follow the approach described in
Ref.~\cite{Grundmann1995}. The electron mass is taken to be
$m_\textrm{e}=$(0.067+0.083$x^\textrm{Al}$)$m_0$, and heavy-hole
anisotropic masses are taken to be
$m_\textrm{hh,z}=$(0.33+0.18$x^\textrm{Al}$)$m_0$,
$m_\textrm{hh,xy}=$(0.11+0.10$x^\textrm{Al}$)$m_0$, where $m_0$ is
the free electron mass. The energy discontinuities at the
GaAs/Al$_{x}$Ga$_{1-x}$As interface are taken to be
0.79$x^\textrm{Al}$~eV for the conduction band and
0.51$x^\textrm{Al}$~eV for the valence band respectively. We solve
single-particle equations for the electron and the hole separately
and the contribution of the Coulomb interaction to the exciton
energy is calculated as a perturbation.

\begin{figure}[h]
\includegraphics[bb=18pt 18pt 345pt 245pt, width=14cm]{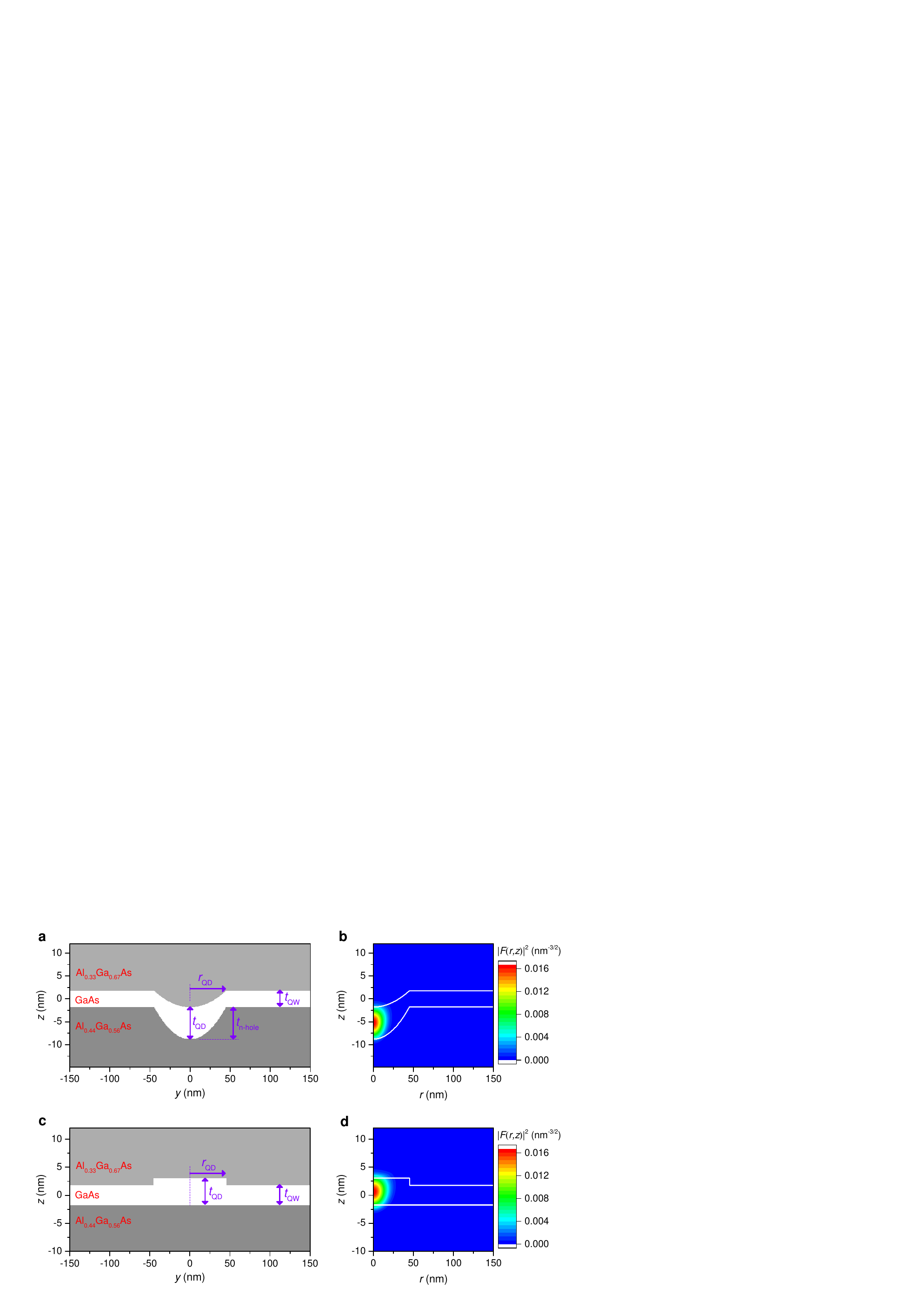}
\caption{\textbf{Model quantum dot structures and calculated
electron envelope wavefunctions.} \textbf{a,} Cross-section of a
model structure of a nanohole GaAs/GaAs quantum dot. Cylindrical
symmetry over $z$ axis is assumed. The geometry of the structure
is determined by quantum well thickness ($t_\textrm{QW}$), quantum
dot thickness ($t_\textrm{QD}$), nanohole depth
($t_\textrm{n-hole}$), and quantum dot radius ($r_\textrm{QD}$).
\textbf{b,} Calculated electron wavefunction profile in
cylindrical coordinates for the dot structure in (\textbf{a}).
White lines show the GaAs/AlGaAs boundaries. \textbf{c,}
Cross-section of a disk-shaped thickness fluctuation quantum dot.
\textbf{d,} Calculated electron wavefunction profile in
cylindrical coordinates for the dot structure in (\textbf{c}).}
\label{Fig:FigSuppStructWF}
\end{figure}

Quantum dots formed by nanohole etching and infilling are modeled
using the structure with a cross-section shown in Supplementary
Fig.~\ref{Fig:FigSuppStructWF}a. We assume cylindric symmetry
which simplifies the problem. The aluminium molar fractions in the
barriers and the thickness of the GaAs quantum well (QW) are taken
according to the growth protocol. The calculated QW exciton
transition energy is found to match the experimental value of
$\sim1.665$~eV (see Supplementary Fig.~\ref{Fig:FigSuppDNPPLE}a)
for $t_\textrm{QW}=3.55$~nm in very good agreement with the design
QW thickness of $t_\textrm{QW}=3.5$~nm. The depth of the nanohole
($t_\textrm{n-hole}$=7.0~nm) the radius of the dot
($r_\textrm{QD}$=45.0~nm), and the thickness of the dot
($t_\textrm{QD}$=7.0~nm) are taken to be comparable to the results
of the AFM studies on similar structures \cite{Atkinson2012}. The
electron wavefunction calculated for such structure is shown in
Supplementary Fig.~\ref{Fig:FigSuppStructWF}b, the optical
transition energy is found to be $\sim$1.585~eV in good agreement
with the emission energies of the long-wavelength (type A) dots.

Short-wavelength quantum dots (type B) were previously shown to
arise from the irregularities in GaAs layer thickness at the rims
of the nanoholes \cite{Ulhaq}. As a simple approximation we model
such dots as disk-shaped QW thickness fluctuations as shown in
Supplementary Fig.~\ref{Fig:FigSuppStructWF}c. With
$r_\textrm{QD}$=45.0~nm and $t_\textrm{QD}$=4.81~nm we find
transition energy of $\sim$1.625~eV in good agreement with
experiment. The corresponding electron wavefunction profile is
shown in Supplementary Fig.~\ref{Fig:FigSuppStructWF}d.

\begin{figure}[h]
\includegraphics[bb=50pt 13pt 405pt 419pt, width=10cm]{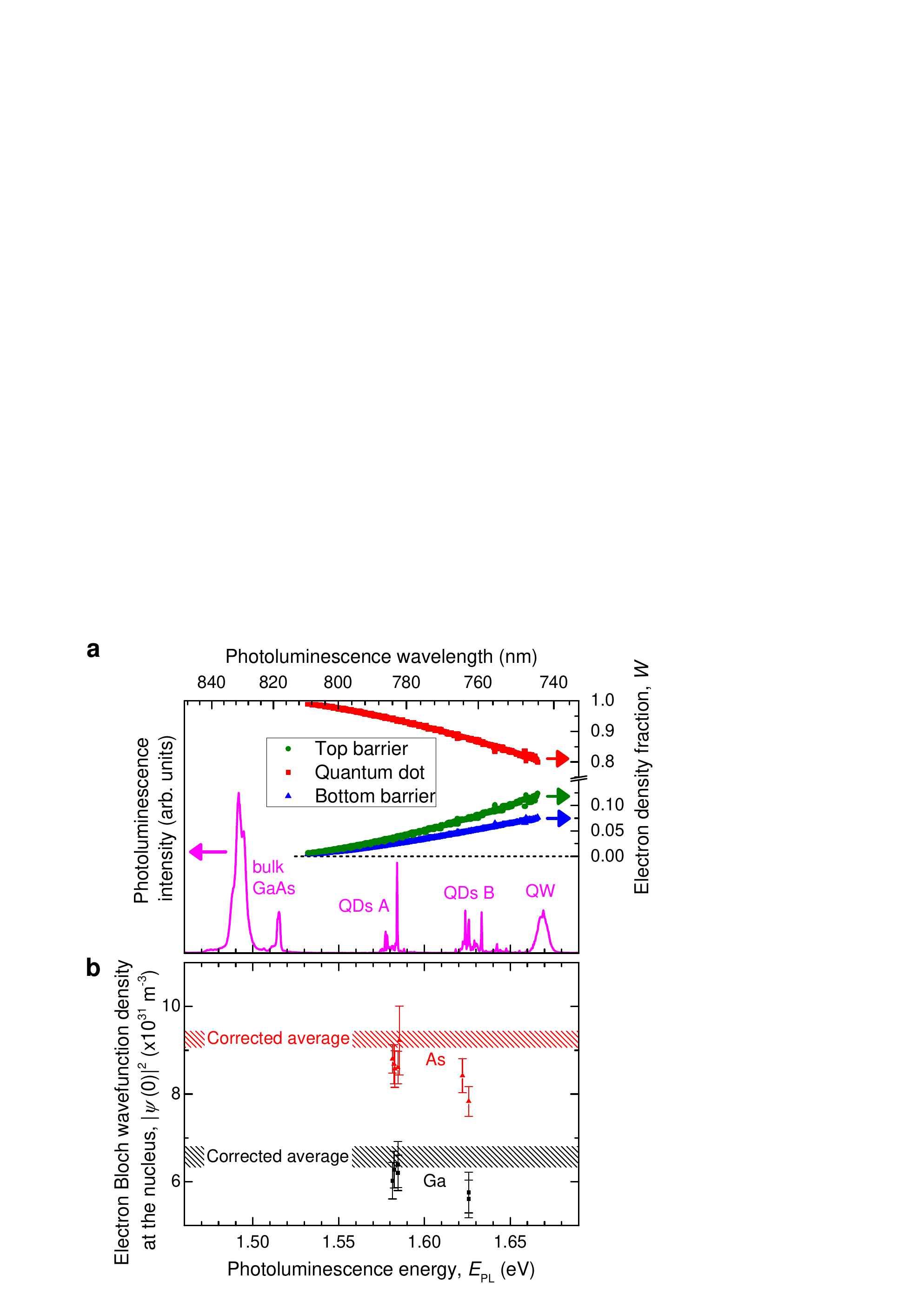}
\caption{\textbf{Electron wavefunction in the nano-hole
GaAs/AlGaAs quantum dots.} \textbf{a,} Typical experimental
photoluminescence spectrum of the studied sample (left scale) and
the fraction of the electron envelope wavefunction density in the
QD and barrier layers as a function of the optical transition
energy (right scale) derived from model calculations. \textbf{b,}
Electron Bloch wavefunction density at the nucleus $|\psi(0)|^2$.
Each symbol corresponds to $|\psi(0)|^2$ derived from experimental
selective rf depolarization of the nuclear spins in an individual
quantum dot. The $|\psi(0)|^2$ values are derived from Eq.~4 of
the main text based on the hyperfine constants derived in turn
from the fitting of the experimental results such as shown in
Figs. 2a, b of the main text. In such fits we assume uniform
nuclear polarization in quantum dots
$P_\textrm{N}^{i}(\mathbf{r}_j)=$const. The results are shown for
Ga (squares) and As (triangles) nuclei with respect to the ground
state exciton luminescence of each quantum dot $E_\textrm{PL}$
with error bars corresponding to 95\% confidence level. By taking
into account the non-uniformity of
$P_\textrm{N}^{i}(\mathbf{r}_j)$ as described in \ref{SI:PNInhom},
we calculate the corrected
$|\psi_\textrm{As}(0)|^2$=(9.25$\pm$0.20)$\times$10$^{31}$~m$^{-3}$
and
$|\psi_\textrm{Ga}(0)|^2$=(6.57$\pm$0.25)$\times$10$^{31}$~m$^{-3}$
averaged over all studied dots as shown by dashed areas
representing  95\% confidence level estimates.}
\label{Fig:FigSuppWFDensity}
\end{figure}

We have performed calculations for a wide range of quantum dot
dimensions $t_\textrm{QD}$, $r_\textrm{QD}$, $t_\textrm{n-hole}$.
As expected, we find that the same $E_\textrm{PL}$ can be obtained
for an infinite number of different combinations of
$t_\textrm{QD}$, $r_\textrm{QD}$, $t_\textrm{n-hole}$: the exciton
optical transition energy alone does not reveal the entire quantum
dot structure. On the other hand, as we show below, the precise
knowledge of the electron wavefunction $F(r,z)$ is not required
for the calculations of the hyperfine shifts (based on
Supplementary Eq.~\ref{eq:Ehfi}). It is sufficient to know the
integral properties of $F(r,z)$ such as the fractions of the
wavefunction density within the quantum dot GaAs layer and the
AlGaAs barriers. These wavefunction density fractions $W$ are
shown in Supplementary Fig.~\ref{Fig:FigSuppWFDensity}a by the
symbols as a function of the exciton transition energy
$E_\textrm{PL}$ calculated for a large number of model quantum dot
structures with different dimensions (both nanohole and
disk-shaped dots are included). It can be seen that the calculated
points reveal clear $W(E_\textrm{PL})$ dependencies: thus using
the experimental $E_\textrm{PL}$ energy derived from a PL spectrum
(such as shown in Supplementary Fig.~\ref{Fig:FigSuppWFDensity}a
by the line) and the calculated $W(E_\textrm{PL})$ one can
estimate the wavefunction fractions $W$ in the QD and barrier
layers for a given studied quantum dot. Since the $W$ values are
the functions of $E_\textrm{PL}$ only, it is not required to know
the exact QD shape and size, instead it is sufficient to choose
any QD model structure that yields $E_\textrm{PL}$ matching the
experimental value. In the following analysis we use the
particular dot model structures of Supplementary
Fig.~\ref{Fig:FigSuppStructWF}a and \ref{Fig:FigSuppStructWF}c
whose $E_\textrm{PL}$ fit the experimentally observed values of
the long- and short-wavelength dots respectively.

\subsection{The role of the nuclear spin polarization inhomogeneity $\beta^i(\mathbf{r}_j)$.\label{SI:PNInhom}}

With $x^i(\mathbf{r}_j)$ and $F(\mathbf{r}_j)$ estimated above, it
is the spatial distribution of the polarization degree
$P_\textrm{N}^{i}(\mathbf{r}_j)$ [or equivalently the distribution
of the inverse nuclear spin temperature $\beta^{i}(\mathbf{r}_j)$]
that needs to be found in order to be able to use Supplementary
Eq.~\ref{eq:Ehfi} to calculate the measured hyperfine shifts. Due
to the complex electron-nuclear spin dynamics, experimental
measurement or the first principle modeling of
$P_\textrm{N}^{i}(\mathbf{r}_j)$ distribution in a quantum dot is
far beyond what can be achieved at present. Yet, as we now show,
it is possible to construct a model for
$P_\textrm{N}^{i}(\mathbf{r}_j)$ that is sufficiently good to
derive electron hyperfine constants $A$ from experimental data.

We start by noting that $P_\textrm{N}^{i}(\mathbf{r}_j)$ should
reach its maximum near the center of the dot
($|\mathbf{r}|\approx0$) where the electron density peaks and the
probability of the electron-nuclear spin flip-flop is maximized.
With increasing $|\mathbf{r}|$ the polarization
$P_\textrm{N}^{i}(\mathbf{r}_j)$ should decay monotonically
towards 0, since the nuclei remain unpolarized away from the dot.
Now let us suppose that the nuclear spins are initially
unpolarized in the entire sample ($P_\textrm{N}\approx0$) and that
the optical cooling is introduced at time $t=0$. At small $t$ the
resulting $P_\textrm{N}^{i}(\mathbf{r}_j)$ will be proportional to
the nuclear spin cooling rate at each point $\mathbf{r}_j$. This
rate in turn is proportional to the electron envelope wavefunction
density $|F(\mathbf{r}_j)|^2$ controlling the electron-nuclear
flip-flop rate, hence
$P_\textrm{N}^{i}(\mathbf{r}_j)\propto|F(\mathbf{r}_j)|^2$ is
expected for short $t$. At longer times $t$, nuclear spin
diffusion \cite{Paget1982,Eberhardt07} will act to establish a
more uniform spatial distribution of
$P_\textrm{N}^{i}(\mathbf{r}_j)$. If the longitudinal nuclear spin
relaxation was absent, spin diffusion would eventually generate
uniform $P_\textrm{N}^{i}=$const independent of $\mathbf{r}_j$ for
$t\rightarrow\infty$. In the real quantum dots in the studied
sample the longitudinal relaxation times are long ($T_1>$500~s)
but not infinite \cite{Ulhaq}. We thus conclude that the real
$P_\textrm{N}^{i}(\mathbf{r}_j)$ produced by optical cooling in
the studied GaAs/AlGaAs quantum dots is between the two limiting
cases of $P_\textrm{N}^{i}(\mathbf{r}_j)=$const and
$P_\textrm{N}^{i}(\mathbf{r}_j)\propto|F(\mathbf{r}_j)|^2$.

\begin{figure}
\includegraphics[bb=9pt 61pt 537pt 699pt, width=10.5cm]{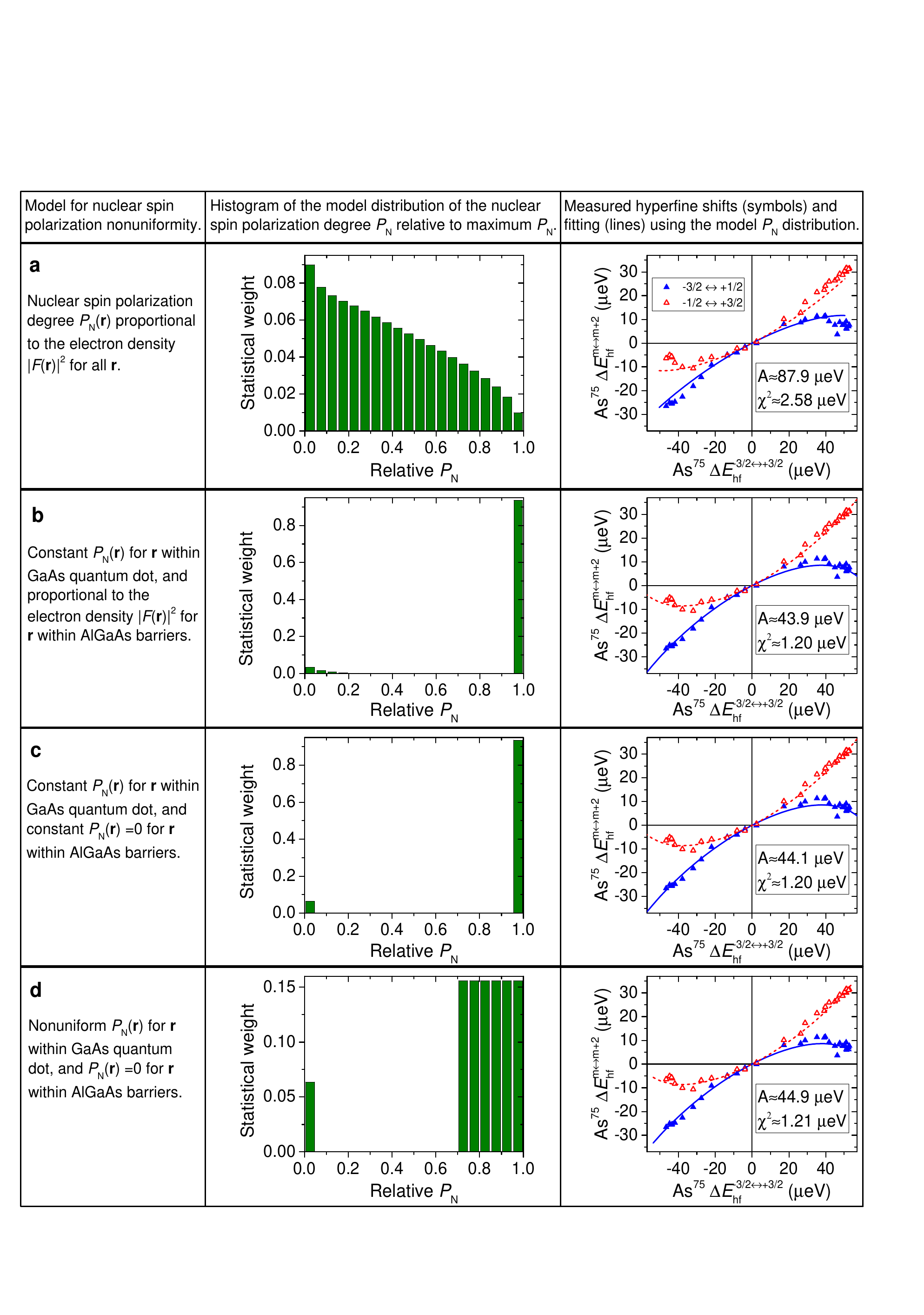}
\caption{\textbf{Derivation of the nuclear spin temperature and
hyperfine constants taking into account spatially inhomogeneous
nuclear spin polarization $P_\textrm{N}$.} Rows \textbf{a-d} show
results for different model distributions of $P_\textrm{N}^{i}$
(left column). Middle column: histograms of the $P_\textrm{N}^{i}$
distributions within the electron wavefunction volume. Right
column: hyperfine shifts $\Delta E_\textrm{hf}^{m\leftrightarrow
m+2}$ induced by selective saturation of two out of the three
dipolar NMR transitions of the spin-3/2 $^{75}$As nuclei as a
function of the hyperfine shift $\Delta
E_\textrm{hf}^{-I\leftrightarrow +I}$ resulting from simultaneous
saturation of all NMR transitions. Symbols show experiments on an
individual quantum dot A1, while lines show model calculations
using Supplementary Equation~\ref{eq:Ehfi} and Eqs.~5 of the main
text. In calculations the molar fraction of As is
$x^\textrm{As}$=1 and the electron envelope wavefunction
$F(\mathbf{r}_j)$ is from Supplementary
Fig.~\ref{Fig:FigSuppStructWF}b. Hyperfine constant $A$ is used as
the only fitting parameter and is shown for each calculation
together with the root mean square fitting residual $\chi^2$.}
\label{Fig:FigSuppPNDistrFit}
\end{figure}

Let us first consider a uniform nuclear spin polarization
$P_\textrm{N}^{i}(\mathbf{r}_j)=$const. In this case, the sum in
the Supplementary Eq.~\ref{eq:Ehfi} becomes a constant $k$
determined only by the structure of the quantum dot, in particular
$k=1$ if $^{75}$As nuclei ($x=\rho=1$) are considered. Under these
assumptions we can use Eqs.~5 of the main text to fit the
experimental data and derive the nuclear spin temperatures and
hyperfine constants $A^i$. The densities of the electron Bloch
wavefunction at the nucleus $|\psi(0)|^2$ calculated from the
fitted values of $A^i$ (see Eq.~4 of the main text) are shown in
Supplementary Fig.~\ref{Fig:FigSuppWFDensity}b by the symbols for
all studied quantum dots against their photoluminescence energy
$E_\textrm{PL}$. For all studied quantum dots we find that the
assumption of $P_\textrm{N}^{i}(\mathbf{r}_j)=$const leads to good
agreement between experiment and fitting, such as shown in Figs.
2a, b of the main text. On the other hand, there is a small but
distinct difference in the fitted $|\psi(0)|^2$ values between the
long-wavelength (type A) and short-wavelength (type B) quantum
dots. As it follows from Supplementary
Fig.~\ref{Fig:FigSuppWFDensity}b, the experiments on type B dots
that have larger fraction of the electron density in the barriers,
give underestimated values of $A^i$ and $|\psi(0)|^2$. This is a
clear sign that the $P_\textrm{N}^{i}(\mathbf{r}_j)=$const
approximation is not exact, and that the deviation arises from the
reduced polarization degree $|P_\textrm{N}^{i}|$ in the barriers.

We now examine the opposite case of the largest possible
inhomogeneity of the nuclear spin polarization. For this we
consider the model structure of Supplementary
Fig.~\ref{Fig:FigSuppStructWF}a with electron envelope
wavefunction shown in Supplementary
Fig.~\ref{Fig:FigSuppStructWF}b, and substitute
$P_\textrm{N}^{i}(\mathbf{r}_j)\propto|F(\mathbf{r}_j)|^2$ in
Supplementary Eq.~\ref{eq:Ehfi} allowing the hyperfine shifts to
be calculated. In order to make the analysis more intuitive we
build a histogram of the weighted $P_\textrm{N}^{i}$ values that
appear in the sum of the Supplementary Eq.~\ref{eq:Ehfi}. For the
case of $P_\textrm{N}^{i}(\mathbf{r}_j)\propto|F(\mathbf{r}_j)|^2$
such histogram is shown in the middle column of the Supplementary
Fig.~\ref{Fig:FigSuppPNDistrFit}a for the As nuclei. Each value on
the horizontal axis is the polarization degree $P_\textrm{N}^{i}$
normalized by its maximum value at the center of the dot, and the
height of each bar reflects the fraction of the nuclei with such
$P_\textrm{N}^{i}$ weighted by the envelope wavefunction density
$|F(\mathbf{r}_j)|^2$ and the molar fraction $x(\mathbf{r}_j)$ at
such nuclear sites. The right graph of the Supplementary
Fig.~\ref{Fig:FigSuppPNDistrFit}a shows experimental (symbols) and
fitted (lines) dependencies $\Delta
E_\textrm{hf}^{m\leftrightarrow m+2}(\Delta
E_\textrm{hf}^{-I\leftrightarrow +I})$ for As nuclei in QD A1. The
fitting yields an unrealistically large
$A^\textrm{As}\approx$87.9~$\mu$eV with a large RMS fitting
residual of $\chi\approx2.58$~$\mu$eV exceeding the experimental
error. Thus we can rule out the
$P_\textrm{N}^{i}(\mathbf{r}_j)\propto|F(\mathbf{r}_j)|^2$ case.

Combining the above observations, we conclude that the real
profile of the nuclear spin polarization degree
$P_\textrm{N}^{i}(\mathbf{r}_j)$ is much closer to the limiting
case of a constant value, rather than to the opposite limit of
strongly inhomogeneous
$P_\textrm{N}^{i}(\mathbf{r}_j)\propto|F(\mathbf{r}_j)|^2$. In
other words, the spatial width of the
$P_\textrm{N}^{i}(\mathbf{r}_j)$ distribution is significantly
larger than that of the envelope wavefunction density
$|F(\mathbf{r}_j)|^2$. On the other hand,
$P_\textrm{N}^{i}(\mathbf{r}_j)$ is not exactly constant, most
likely due to the reduced $|P_\textrm{N}^{i}|$ in the AlGaAs
barriers. We now discuss how this residual spatial inhomogeneity
of $P_\textrm{N}^{i}(\mathbf{r}_j)$ can be accounted for in order
to improve the accuracy of the hyperfine constant measurement.

Let us assume that optical cooling produces constant
$P_\textrm{N}^{i}$ within the quantum dot layer, while outside the
dot the polarization degree scales as
$P_\textrm{N}^{i}(\mathbf{r}_j)\propto|F(\mathbf{r}_j)|^2$. The
corresponding histogram of $P_\textrm{N}^{i}$ and fitted $\Delta
E_\textrm{hf}^{m\leftrightarrow m+2}(\Delta
E_\textrm{hf}^{-I\leftrightarrow +I})$ dependencies are shown in
Supplementary Fig.~\ref{Fig:FigSuppPNDistrFit}b. Such a model for
$P_\textrm{N}^{i}(\mathbf{r}_j)$ gives an accurate fit with an RMS
residual of $\chi\approx1.20$~$\mu$eV within the experimental
error of the electron hyperfine shift measurements. The assumption
of a constant level of $P_\textrm{N}^{i}$ within the dot volume
can be well justified: long optical cooling times ($>10$~s) used
in our experiments give sufficient time for nuclear spin
polarization to be redistributed via spin diffusion. On the other
hand the reduction of $P_\textrm{N}^{i}$ in AlGaAs barriers can be
understood to arise from the quadrupolar induced suppression of
the spin diffusion at the GaAs/AlGaAs interfaces
\cite{Nikolaenko2009,Ulhaq}.

Very similar fitted value of the hyperfine constant $A$ is
obtained if we assume a simple bimodal distribution for
$P_\textrm{N}^{i}$ (constant $P_\textrm{N}^{i}$ within the dot and
$P_\textrm{N}^{i}=0$ in the barriers as shown in Supplementary
Fig.~\ref{Fig:FigSuppPNDistrFit}c). It is thus evident that the
detailed form of $P_\textrm{N}^{i}(\mathbf{r}_j)$ distribution in
the barriers is not critical due to the small overall effect of
the barrier nuclear spin polarization. Importantly, when we
perform fitting with bimodal distributions of Supplementary
Figs.~\ref{Fig:FigSuppPNDistrFit}b,c we obtain very close values
of the hyperfine constant $A$ (and hence $|\psi(0)|^2$) for both
type A and type B dots -- this is a good indication that bimodal
distribution of $P_\textrm{N}^{i}$ is a good approximation to the
real distribution of the optically induced
$P_\textrm{N}^{i}(\mathbf{r}_j)$.

We now note that the fitting with a bimodal distribution of
Supplementary Fig.~\ref{Fig:FigSuppPNDistrFit}c is equivalent to
fitting with a constant $P_\textrm{N}^{i}(\mathbf{r}_j)$ in the
entire sample, but with hyperfine constant $A$ replaced by $kA$.
This is because the nuclei with $P_\textrm{N}^{i}=0$ do not
contribute to the hyperfine shifts $\Delta
E_\textrm{hf}^{m\leftrightarrow m+2}$ and $\Delta
E_\textrm{hf}^{-I\leftrightarrow +I}$, in which case the
additional factor $k$ equals $W$, where $W$ is the fraction of the
electron density in the GaAs QD layer shown in Supplementary
Fig.~\ref{Fig:FigSuppWFDensity}a (we consider here the case of
$^{75}$As where $\rho=x=1$).

Finally, we examine a case where the barrier nuclei are not
polarized ($P_\textrm{N}^{i}=0$), while the polarization of the QD
nuclei is not constant. As an example we use a rectangular
distribution with a histogram shown in the middle plot of
Supplementary Fig.~\ref{Fig:FigSuppPNDistrFit}d, where we allow
the weighted $P_\textrm{N}^{i}$ to be uniformly spread between
70\% and 100\% of its maximum value, which is likely an
exaggeration of the inhomogeneity in a real quantum dot. As the
right plot of Supplementary Fig.~\ref{Fig:FigSuppPNDistrFit}d
shows we still find a very good fit with fitted hyperfine constant
$A$ similar to that obtained from a bimodal distributions of
Supplementary Figs.~\ref{Fig:FigSuppPNDistrFit}b, c. We also find
very similar average polarization degrees $P_\textrm{N}^{i}$
derived from the fits with different $P_\textrm{N}^{i}$
distributions shown in Supplementary
Figs.~\ref{Fig:FigSuppPNDistrFit}b-d.

We thus summarize with the following conclusions. From the
measurements on short- and long-wavelength quantum dots we
conclude that $P_\textrm{N}^{i}$ is reduced in the barriers, which
is explained by the suppression of the nuclear spin diffusion at
the GaAs/AlGaAs interfaces \cite{Nikolaenko2009,Ulhaq}. Our
measurement technique not very sensitive to the details of the
spatial distribution $P_\textrm{N}^{i}(\mathbf{r}_j)$ of the
nuclear spin polarization. This however, comes as an advantage,
allowing robust measurement of the average $P_\textrm{N}^{i}$ and
$T_\textrm{N}$ within the GaAs QD layer regardless of the details
of the $P_\textrm{N}^{i}(\mathbf{r}_j)$ profile. In this
Supplementary Note we have presented a detailed first-principles
procedure for deriving the electron hyperfine constants $A$ from
the NMR experiments taking into account the spatial inhomogeneity
of the nuclear spin polarization degree
$P_\textrm{N}^{i}(\mathbf{r}_j)$. At the same time we have shown
that a simplified analysis assuming
$P_\textrm{N}^{i}(\mathbf{r}_j)=$const (presented in the main
text) gives very similar results as long as
$P_\textrm{N}^{i}(\mathbf{r}_j)$ distribution satisfies rather
generic constraints. The only difference is that instead of the
hyperfine constants $A$, the simplified model fitting yields a
scaled $kA$ where the structural factor $k\leq1$ depends on the
electron wavefunction density fraction $W$ within the GaAs quantum
dot volume. This fraction $W$ can be estimated by solving the
Schrodinger equation for \textit{any} reasonable model structure
whose photoluminescence energy $E_\textrm{PL}$ matches the
experimentally measured $E_\textrm{PL}$ -- it is not required to
know the exact shape and size of the quantum dot. Even if the
value of $k$ can not be estimated, the simplified analysis still
gives a reliable measure of the average $P_\textrm{N}^{i}$ within
the dot volume, making the techniques reported here a valuable
tool for analysis of the nuclear spin bath thermodynamics in
semiconductor quantum dots.

\end{document}